\documentclass[10pt]{article}
\usepackage{fullpage}
\usepackage{amsmath}
\usepackage{amssymb}
\usepackage{amsfonts}
\usepackage[dvips]{epsfig}
\usepackage{color}
\usepackage{cite}

\def\##1{{\underline{#1}}}
\def\=#1{\underline{\underline{#1}}}

\def\+#1{\underline{\bf #1}}
\def\*#1{\underline{\underline{\bf #1}}}

\def\r#1{(\ref{#1})}
\def\l#1{\label{#1}}
\def\c#1{\cite{#1}}

\def\le{\left(}
\def\ri{\right)}
\def\les{\left[}
\def\ris{\right]}
\def\lec{\left\{}
\def\ric{\right\}}

\def\.{\mbox{ \tiny{$^\bullet$} }}

\def\eps{\varepsilon}

\def\epso{\eps_{\scriptscriptstyle 0}}
\def\lambdao{\lambda_{\scriptscriptstyle 0}}
\def\muo{\mu_{\scriptscriptstyle 0}}

\def\ko{k_{\scriptscriptstyle 0}}

\def\ux{\hat{\#u}_{\,x}}
\def\uy{\hat{\#u}_{\,y}}
\def\uz{\hat{\#u}_{\,z}}

\def\calA{{\cal A}}
\def\calB{{\cal B}}

\begin{document}

\begin{center}

\LARGE{ {\bf Electromagnetic surface waves guided by the planar interface of isotropic chiral materials
}}
\end{center}
\begin{center}
\vspace{10mm} \large

{\bf Maimoona Naheed}$^1$, {\bf Muhammad Faryad}$^2$ and {\bf Tom G. Mackay}$^{3,4,}$\footnote{Corresponding author. E--mail: T.Mackay@ed.ac.uk}

\vspace{2mm}

$^1$\emph{Department of Electronics, Quaid-i-Azam University, Islamabad, Pakistan}

\vspace{2mm}

$^2$\emph{Department of Physics, Lahore University of Management Sciences, Lahore, Pakistan}

\vspace{2mm}

$^3$\emph{School of Mathematics  and
   Maxwell Institute for Mathematical Sciences, \\University of Edinburgh, Edinburgh
EH9 3JZ, United Kingdom}

 \vspace{2mm}

 $^4$\emph{NanoMM---Nanoengineered Metamaterials Group, Department of Engineering Science and
Mechanics, Pennsylvania State University, University Park, PA
16802--6812, USA}

\vspace{5mm}
\normalsize

\end{center}

\begin{center}
\vspace{15mm} {\bf Abstract}
\end{center}

\vspace{5mm}

The propagation of electromagnetic surface waves guided by the planar interface of two isotropic chiral materials, namely materials $\calA$ and $\calB$, was investigated by numerically solving the associated canonical boundary-value problem. Isotropic chiral material $\calB$  was modeled as a homogenized composite material,  arising from the homogenization of an isotropic chiral component material and an isotropic achiral, nonmagnetic,  component material characterized by the relative permittivity $\eps_a^\calB$. 
 Changes in the nature of the surface waves were explored as the volume fraction $f_a^\calB$ of the  achiral component material varied. 
Surface waves are supported only for certain ranges of $f_a^\calB$; within these ranges  only one surface wave, characterized by its relative wavenumber $q$, is supported at each value of $f_a^\calB$.
For $\mbox{Re} \lec \eps_a^\calB \ric > 0 $,
as $\left| \mbox{Im} \lec \eps_a^\calB \ric \right|$ increases
 surface waves are supported for larger ranges of $f_a^\calB$  and $\left| \mbox{Im} \lec q \ric \right|$  for these surface waves increases.
 For $\mbox{Re} \lec \eps_a^\calB \ric < 0 $, 
 as $ \mbox{Im} \lec \eps_a^\calB \ric $ increases
 the ranges of $f_a^\calB$  that support surface-wave propagation are almost unchanged but  $ \mbox{Im} \lec q \ric $  for these surface waves decreases.
 The surface waves supported when $\mbox{Re} \lec \eps_a^\calB \ric < 0 $ may be regarded as akin to surface-plasmon-polariton waves, but those supported for
when $\mbox{Re} \lec \eps_a^\calB \ric > 0 $ may not. \\

\vspace{5mm}

\section{Introduction}

Electromagnetic surface waves are guided by 
the planar interface of two dissimilar materials. Various  types of electromagnetic surface wave have been identified since the early 1900s right up to the present day. 
The type assigned to a given surface wave depends upon whether the partnering materials are isotropic or anisotropic, dissipative or nondissipative, homogeneous or nonhomogeneous,  etc. \c{ESW_book}. The planar interface of a plasmonic material
and a dielectric material guides the surface-plasmon-polariton (SPP) wave \c{Pitarke,Maier}, which is  the most well-known type of electromagnetic surface wave.
The partnering materials for SPP waves may be  isotropic or anisotropic \c{Jacob}.   Uses for 
these
waves are found in  optical sensing  \c{Homola_book,AZLe}.  Another well-known type of electromagnetic surface wave
is the Dyakonov wave \c{Marchevskii,Dyakonov88}, which is guided by the planar interface of an isotropic dielectric material and an anisotropic dielectric material \c{Walker98,Takayama_exp}. Dyakonov waves  are
generally associated with small angular existence domains \c{DSWreview} but larger angular existence domains can be supported if the partnering materials are dissipative \cite{MLieee,FFespo}.
Surface waves that are intermediate in character,  in part akin to  SPP waves and in part akin to Dyakonov waves, can be supported by hyperbolic materials \c{Jacob,Takayama2,Li}.

Owing to their inherent magneto-electric coupling,
chiral materials \c{Beltrami}  offer wider opportunities for surface-wave propagation than  achiral materials. To date, there have been relatively 
 few studies of the surface waves supported by  chiral materials \c{Pattanayak,Engheta,Fantino,Pellegrini}, as compared to achiral materials.  Most of these  studies have concentrated on interfaces of
 nondissipative  chiral materials and  isotropic plasmonic materials. The corresponding surface waves in these studies are akin to SPP waves.
 Recently, surface waves guided by the planar interfaces of  chiral materials and  anisotropic achiral materials were explored numerically \c{Noonan}. In this study  surface waves akin to Dyakonov waves were  found to be supported when the achiral partnering material was a dielectric material while surface waves akin  to SPP waves 
 were  found to be supported when the achiral partnering material was a plasmonic material. 
 
In the following we consider the previously-unexplored case of surface waves guided by the planar interface of two isotropic chiral materials. The dispersion relation corresponding to the canonical boundary-value problem  \c{ESW_book}  is derived and numerical solutions are extracted. To allow greater flexibility for our numerical investigations, both
isotropic chiral partnering materials are modeled as homogenized composite materials (HCMs) \c{MAEH}.
Furthermore,
 a novel type of chiral material that can simultaneously support attenuation and amplification of plane waves \c{ML_JO}, depending upon circular polarization state, is utilized.

 As regards notation,
 the permittivity and permeability of free space are written as $\epso$ and $\muo$, respectively. 
 The free-space wavelength is $\lambdao = 2 \pi / \ko$ with
 the free-space wavenumber being $\ko = \omega \sqrt{\epso \muo}$ and  $\omega$ being the angular frequency. The operators $\mbox{Re} \lec \. \ric$ 
and $\mbox{Im} \lec \. \ric$ deliver the
real and imaginary parts of  complex-valued quantities, and $i = \sqrt{-1}$. 
Single underlining signifies a 3 vector 
and the triad of unit vectors aligned with the Cartesian axes are   $\lec \ux, \uy, \uz \ric$. Matrixes are enclosed by square brackets.

\section{Theory: canonical boundary-value problem for surface-wave propagation}

Let us consider the canonical boundary-value problem for surface waves  \c{ESW_book} guided by the planar  interface of two different isotropic chiral  materials. Both partnering materials are homogeneous.
The isotropic chiral  material, labeled $\mathcal{A}$, fills
 the half-space $z>0$ and is characterized by the frequency-domain Tellegen constitutive relations \c{Beltrami}
\begin{equation} \l{cr}
\left.
\begin{array}{l}
 \#D (\#r) = \epso \eps_\mathcal{A} \#E (\#r) + i \sqrt{\epso \muo}  \xi_\mathcal{A} \#H (\#r) \,\\ [5pt]
 \#B (\#r)= - i \sqrt{\epso \muo} \xi_\mathcal{A} \#E (\#r) + \muo \mu_\mathcal{A} \#H (\#r) \,
\end{array}
\right\}\,, \qquad z>0 ,
\end{equation}
while
the isotropic chiral  material, labeled $\mathcal{B}$, fills
 the half-space $z<0$ and is characterized by the frequency-domain Tellegen constitutive relations \c{Beltrami}
\begin{equation} \l{cr2}
\left.
\begin{array}{l}
 \#D (\#r) = \epso \eps_\mathcal{B} \#E (\#r) + i \sqrt{\epso \muo}  \xi_\mathcal{B} \#H (\#r) \,\\ [5pt]
 \#B (\#r)= - i \sqrt{\epso \muo} \xi_\mathcal{B} \#E (\#r) + \muo \mu_\mathcal{B} \#H (\#r) \,
\end{array}
\right\}\, , \qquad z<0.
\end{equation}
The relative permittivity scalars $\eps_{\calA,\calB}$, the relative  permeability scalars $\mu_{\calA,\calB}$, and the relative
chirality pseudoscalars
$\xi_{\calA,\calB}$ are frequency dependent and generally complex valued, per the principle
of causality embodied by the Kramers--Kronig relations \cite{2lines}.

The electromagnetic field phasors in the  partnering materials $\calA$ and $\calB$ are represented by
\begin{equation} \label{planewave}
\left.\begin{array}{l}
 \#E_{\,\ell} (\#r)=  \#{\mathcal E}_{\,\ell} \,\exp\left({i\#k_{\,\ell} \cdot\#r}\right) \\[4pt]
\#H_{\,\ell}(\#r)= \#{\mathcal H}_{\,\ell} \,\exp\left({i\#k_{\,\ell} \cdot\#r}\right)
 \end{array}\right\}\,, \qquad   \ell \in\left\{ \calA, \calB \right\}\,.
\end{equation}
The  amplitude vectors $ \#{\mathcal E}_{\,\ell} $ and $ \#{\mathcal H}_{\,\ell}$ have complex-valued components, and so does
the wave vector
 $\#k_{\,\ell}$.  The field phasors (and the wave vector)   can vary with angular frequency $\omega$. Without loss of generality, we consider  surface-wave  propagation parallel to $\ux$ in the $xy$ plane; i.e., $\uy\cdot\#k_{\,\ell}\equiv 0$.


Assuming an $\exp(-i\omega t)$ time-dependence, the Maxwell curl postulates  yield
\begin{equation} \l{MP_A}
\left.
\begin{array}{l}
\#k_{\, \ell} \times \#{\mathcal E}_{\,\ell} - \omega  \le - i \sqrt{\epso \muo} \xi_{\ell}  \#{\mathcal E}_{\,\ell} +
\muo \mu_\ell \#{\mathcal H}_{\,\ell} \ri = \#0
\vspace{4pt} \\
\#k_{\, \ell} \times \#{\mathcal H}_{\,\ell} + \omega \le  \epso \eps_{\ell}  \#{\mathcal E}_{\,\ell}
+ i \sqrt{\epso \muo} \xi_{\ell}  \#{\mathcal H}_{\,\ell}
\ri = \#0
\end{array}
\right\}\,,
\end{equation}
with $\ell = \calA$ for half-space $z>0$ and $\ell = \calB $ for half-space $z<0$. The wave vector
\begin{equation} \l{wvA}
\#k_{\, \ell} \equiv \left\{ \begin{array}{l}
\#k_{\, \calA} = \ko \le q \, \ux + i \alpha_\calA \uz \, \ri, \qquad $z>0$ \vspace{4pt}\\
\#k_{\, \calB} = \ko \le q \, \ux - i \alpha_\calB \uz \, \ri, \qquad $z<0$
\end{array} \right. ,
\end{equation}
with
 ${\rm Re}\lec\alpha_{\ell} \ric>0$   $\le \ell \in\left\{ \calA, \calB \right\} \ri$ for surface-wave propagation.
On combining  Eqs.~\r{MP_A} and Eq.~\r{wvA}, a biquadratic dispersion relation
emerges for $\alpha_\ell$  $\le \ell \in\left\{ \calA, \calB \right\} \ri$. The two   $\alpha_\ell$  roots with non-negative real parts are identified as
\begin{equation} \l{a_decay_const}
\left.
\begin{array}{l}
\alpha_{\ell 1} =  \sqrt{q^2 - \kappa_{\ell R}^2  } \vspace{8pt}\\
\alpha_{\ell 2} =
 \sqrt{q^2 - \kappa_{\ell L}^2  }
\end{array}
\right\}\,,
\end{equation}
with the complex-valued scalars
\begin{equation}
\left.
\begin{array}{l}
\kappa_{\ell R}=  \sqrt{\eps_\ell \mu_\ell } + \xi_\ell  \vspace{6pt}\\
\kappa_{\ell L}=  \sqrt{\eps_\ell \mu_\ell } - \xi_\ell  \end{array} \right\}
\end{equation}
being associated with the relative wavenumbers for right and left circularly-polarized light, respectively,  in an unbounded chiral medium \c{Beltrami}.
Accordingly the field-phasor amplitudes are given as
\begin{equation}
\left.
\begin{array}{l} \l{A_p}
\#{\mathcal E}_{\,\ell} = C_{\ell 1}\, \#{\mathcal E}_{\,\ell 1} + C_{\ell 2} \,\#{\mathcal E}_{\,\ell 2}  \vspace{6pt}\\
\#{\mathcal H}_{\,\ell} = \displaystyle{\sqrt{\frac{\epso}{\muo}} \sqrt{\frac{\eps_\ell}{\mu_\ell}
}  \le C_{\ell 1} \,\#{\mathcal H}_{\,\ell 1} + C_{\ell 2} \,\#{\mathcal H}_{\,\ell 2} \ri  }
\end{array}
\right\}, \quad \ell \in\left\{ \calA, \calB \right\},
\end{equation}
where the vectors
\begin{equation}
\left.
\begin{array}{l}  \l{A_p2}
\#{\mathcal E}_{\,\calA 1} =   \alpha_{\calA 1}  \, \ux
+ \kappa_{\calA R} \, \uy +  i q  \, \uz \vspace{6pt}\\
\#{\mathcal E}_{\,\calA 2} = 
- \alpha_{\calA 2}  \, \ux
+ \kappa_{\calA L} \, \uy -  i q  \, \uz
 \vspace{6pt}\\
\#{\mathcal H}_{\,\calA 1} = \displaystyle{
-i  \alpha_{\calA 1}  \, \ux
- i \kappa_{\calA R} \, \uy +
q  \, \uz
   }
 \vspace{6pt}\\
\#{\mathcal H}_{\,\calA 2} = 
\displaystyle{
-i  \alpha_{\calA 2}  \, \ux +
 i \kappa_{\calA L} \, \uy +
q  \, \uz
   }
\end{array}
\right\}\,
\end{equation}
and
\begin{equation}
\left.
\begin{array}{l}  \l{B_p2}
\#{\mathcal E}_{\,\calB 1} =  - \alpha_{\calB 1}  \, \ux
+ \kappa_{\calB R} \, \uy +  i q  \, \uz \vspace{6pt}\\
\#{\mathcal E}_{\,\calB 2} = 
 \alpha_{\calB 2}  \, \ux
+ \kappa_{\calB L} \, \uy -  i q  \, \uz
 \vspace{6pt}\\
\#{\mathcal H}_{\,\calB 1} = \displaystyle{
 i  \alpha_{\calB 1}  \, \ux
- i \kappa_{\calB R} \, \uy +
q  \, \uz
   }
 \vspace{6pt}\\
\#{\mathcal H}_{\,\calB 2} = 
\displaystyle{
i  \alpha_{\calB 2}  \, \ux +
 i \kappa_{\calB L} \, \uy +
q  \, \uz
   }
\end{array}
\right\}\,.
\end{equation}

 The four scalars 
 $C_{\calA 1,2} $ and $C_{\calB 1,2} $ introduced   in Eqs.~\r{A_p}, as well as
the relative wavenumber $q$,
 are determined by enforcing boundary conditions
 across the planar interface $z=0$, as follows.
 The continuity of tangential  components of the electric and magnetic field
 phasors across the planar interface  $z=0$ imposes
  four conditions \c{Chen}
which may be represented compactly as
\begin{equation} \l{M0}
\les M \ris \. \les \begin{array}{c}
 C_{\calA 1} \\
  C_{\calA 2} \\
   C_{\calB 1} \\
    C_{\calB 2}
\end{array}
 \ris =  \les \begin{array}{c}
 0 \\
  0 \\
   0 \\
    0
\end{array}
 \ris,
\end{equation}
wherein the 4$\times$4 matrix 
\begin{equation} \l{Mdef}
\les M \ris
=
\les
\begin{array}{cccc}
 \alpha_{\calA 1} &  - \alpha_{\calA 2} &  \alpha_{\calB 1}  & - \alpha_{\calB 2}  \\
  \kappa_{\calA R} &  \kappa_{\calA L} & - \kappa_{\calB R}  & -  \kappa_{\calB L} \vspace{8pt} \\
  \displaystyle{ \alpha_{\calA 1} \sqrt{\frac{\eps_\calA}{\mu_\calA}}}  &
 \displaystyle{\alpha_{\calA 2} \sqrt{\frac{\eps_\calA}{\mu_\calA}}}   &
  \displaystyle{\alpha_{\calB 1} \sqrt{\frac{\eps_\calB}{\mu_\calB}}}  &
   \displaystyle{\alpha_{\calB 2} \sqrt{\frac{\eps_\calB}{\mu_\calB}}} \vspace{8pt} \\
     \displaystyle{- \kappa_{\calA R} \sqrt{\frac{\eps_\calA}{\mu_\calA}}}  &
       \displaystyle{ \kappa_{\calA L} \sqrt{\frac{\eps_\calA}{\mu_\calA}}}  &
         \displaystyle{ \kappa_{\calB R} \sqrt{\frac{\eps_\calB}{\mu_\calB}}}  &
           \displaystyle{- \kappa_{\calB L} \sqrt{\frac{\eps_\calB}{\mu_\calB}}}  
   \end{array}  
\ris.
\end{equation}
For a nontrivial solution to Eq.~\r{M0}, the matrix $\les M \ris$ must be singular. Hence the dispersion equation 
\begin{equation} \l{disp_equation}
\det \les M \ris = 0
\end{equation}
arises, from which $q$ may be extracted, generally by numerical means \c{N-R}. Once $q$ is known,  relative values of the 
four
scalars  $C_{\calA 1,2} $ and $C_{\calB 1,2} $ can be determined from Eq.~\r{M0}  by straightforward algebraic manipulations.

\section{Numerical studies}

The partnering materials $\calA$ and $\calB$ are both isotropic chiral materials \c{Beltrami}, per the Tellegen constitutive relations \r{cr}.
To allow flexibility in specifying the constitutive parameters for these materials, both partnering materials $\calA$ and $\calB$ are modeled as HCMs.
Specifically, partnering material $\ell \in \lec \calA, \calB \ric$ arises from the homogenization of two component materials, namely component material $\ell_a$ which is an achiral, nonmagnetic, isotropic material characterized by the relative permittivity
$\eps^\ell_a$ and component material $\ell_b$ which is an isotropic chiral material characterized by the relative constitutive parameters $\eps^\ell_b$, $\mu^\ell_b$ and $\xi^\ell_b$ per the   Tellegen constitutive relations \r{cr}.
Both component materials in each half-space $z \lessgtr 0$  are assumed to be randomly distributed as electrically small spheres, with the volume fraction of  component material $\ell_a$ being $f^\ell_a$ and that of 
 component material $\ell_b$ being $f^\ell_b = 1 - f^\ell_a$.

Estimates of the constitutive parameters of the partnering materials $\calA$ and $\calB$, namely $\eps^\ell, \mu^\ell$ and $\xi^\ell$  $\le \ell \in \lec \calA, \calB \ric \ri$, are provided by the Bruggeman homogenization formalism \c{MAEH,Kampia}. This process involves numerically solving the following nonlinear matrix equation
\begin{eqnarray} \l{Br_eq}
&& f^\ell_a \le \les K^\ell_a \ris - \les K^\ell \ris \ri \. \lec \les I \ris + \les D^\ell \ris \. \le \les K^\ell_a \ris - \les K^\ell \ris  \ri    \ric^{-1} = \nonumber \\
&&   f^\ell_b \le \les K^\ell \ris - \les K^\ell_b \ris \ri \. \lec \les I \ris + \les D^\ell \ris \. \le \les K^\ell_b \ris - \les K^\ell \ris  \ri    \ric^{-1}, \qquad
\end{eqnarray}
wherein
the constitutive  2$\times$2 matrixes
\begin{equation}
\left.
\begin{array}{c}
 \les K^\ell_a \ris  = \les  \begin{array}{cc}  \eps^\ell_a & 0  \\ 0 & 1 \end{array} \ris
 \vspace{4pt}\\
 \les K^\ell_b \ris  = \les  \begin{array}{cc}  \eps^\ell_b & i \xi^\ell_b  \\ - i \xi^\ell_b & \mu^\ell_b \end{array} \ris
 \vspace{4pt}\\
 \les K^\ell \ris  = \les  \begin{array}{cc}  \eps^\ell & i \xi^\ell  \\ - i \xi^\ell & \mu^\ell \end{array} \ris
\end{array}
\right\},
\end{equation}
 the  depolarization  2$\times$2 matrix
\begin{equation}
 \les D^\ell \ris  = \frac{1}{3 \les \eps^\ell \mu^\ell + \le  \xi^\ell \ri^2 \ris}\les  \begin{array}{cc}  \mu^\ell & - \xi^\ell  \\ \xi^\ell & \eps^\ell \end{array} \ris,
\end{equation}
and $\les I \ris$ is red{the  2$\times$2  identity matrix.}

For the isotropic chiral component material, for both partnering materials $\calA$ and $\calB$, we fixed $\eps^\calA_b \equiv \eps^\calB_b = 3 + 0.01i$, 
$\mu^\calA_b \equiv \mu^\calB_b = 0.95 + 0.0002i$,  and $\xi^\calA_b \equiv \xi^\calB_b = 0.1 + 0.001i$, these values being consistent with certain mildly dissipative,  isotropic chiral metamaterials \c{Kwon}.
For the isotropic achiral component material for partnering material $\calA$, we fixed $\eps^\calA_a = 2 - 0.02i $.  Thus, the  component material $\calA_a$ is an active material.
 The selected value  of $\eps^\calA_a$ falls within the range commonly used for active components of metamaterials in the visible regime. For example,
a mixture of  Rhodamine 800  and  Rhodamine 6G,  yields a relative permittivity with
real part in the range $\le 1.8, 2.3 \ri$ and
 imaginary part in the range
$\le -0.15, -0.02 \ri$ for the frequency range 440--500 THz,
depending upon the relative concentrations  and the external pumping rate \c{Sun_APL}.
The volume fraction of component material $\calA_a$ was fixed at $f^\calA_a = 0.3$. Consequently, the Bruggeman equation \r{Br_eq} delivers the 
constitutive parameter 
estimates $\eps^\calA = 2.6721 - 0.0007 i$, $\mu^\calA = 0.9645+0.0001$, and $\xi^\calA = 0.0675+ 0.0006 i$. Hence, partnering material $\calA$ is an
 isotropic chiral material that simultaneously supports amplification and attenuation of plane waves, depending upon the state of circular polarization  \c{ML_JO}.
Several  manifestations of simultaneous amplification and attenuation of plane waves have been reported recently \c{ML_PRA}, including within the context of surface waves \c{ML_SPP_JO,ML_Dyakonov}.

\begin{figure}[!htb]
\centering
\includegraphics[width=5.8cm]{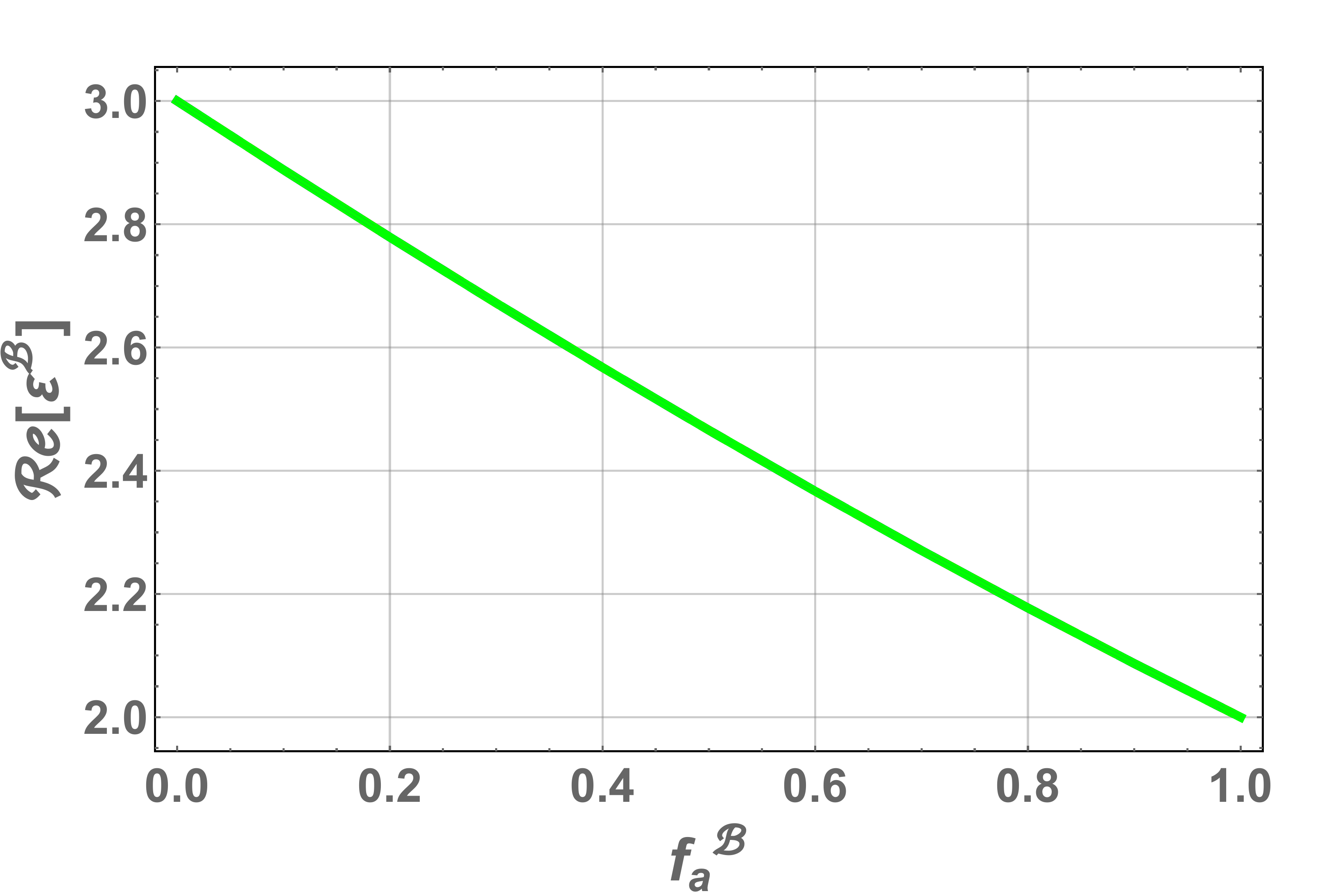} \includegraphics[width=5.8cm]{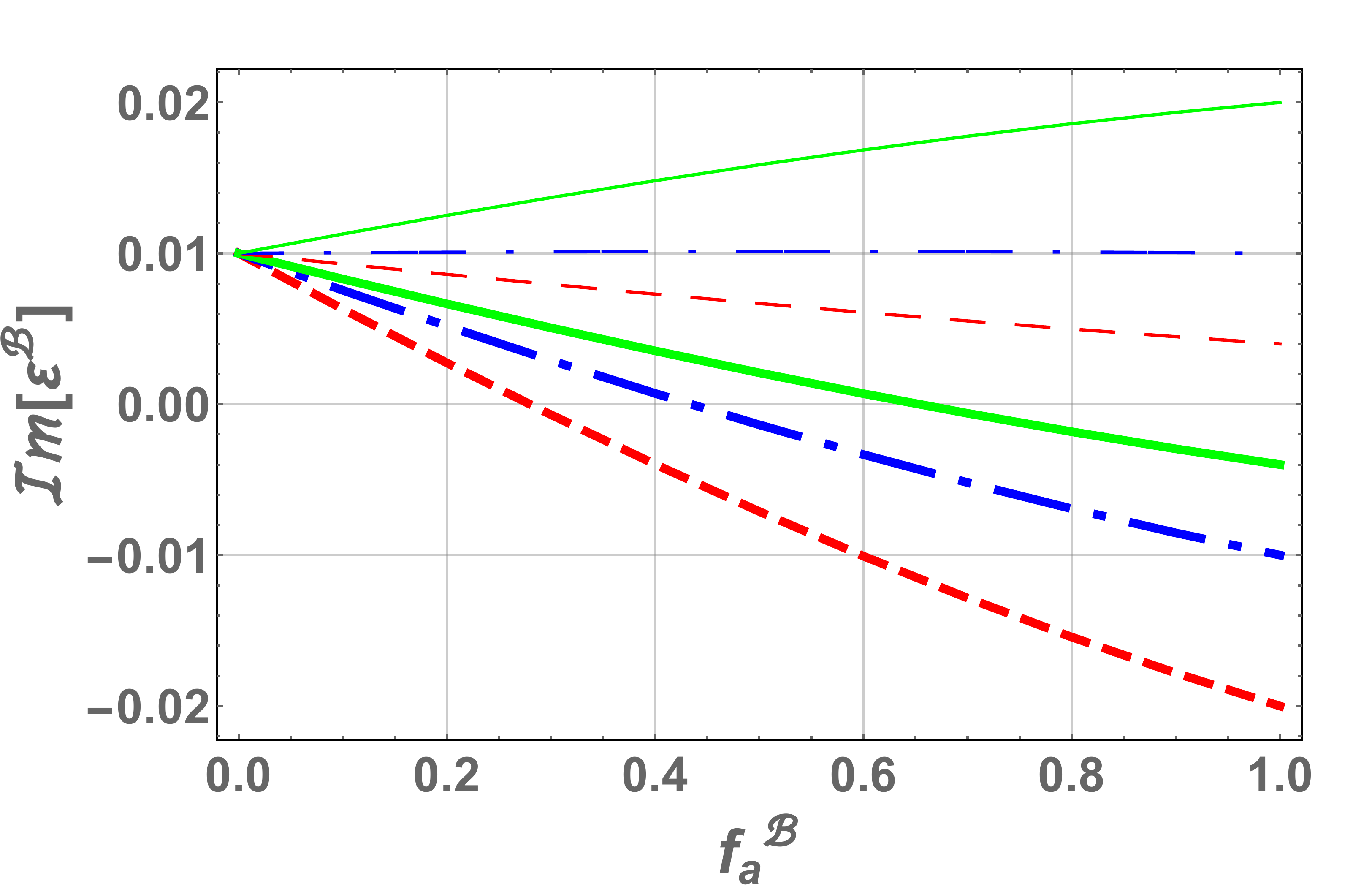}\\
\includegraphics[width=5.8cm]{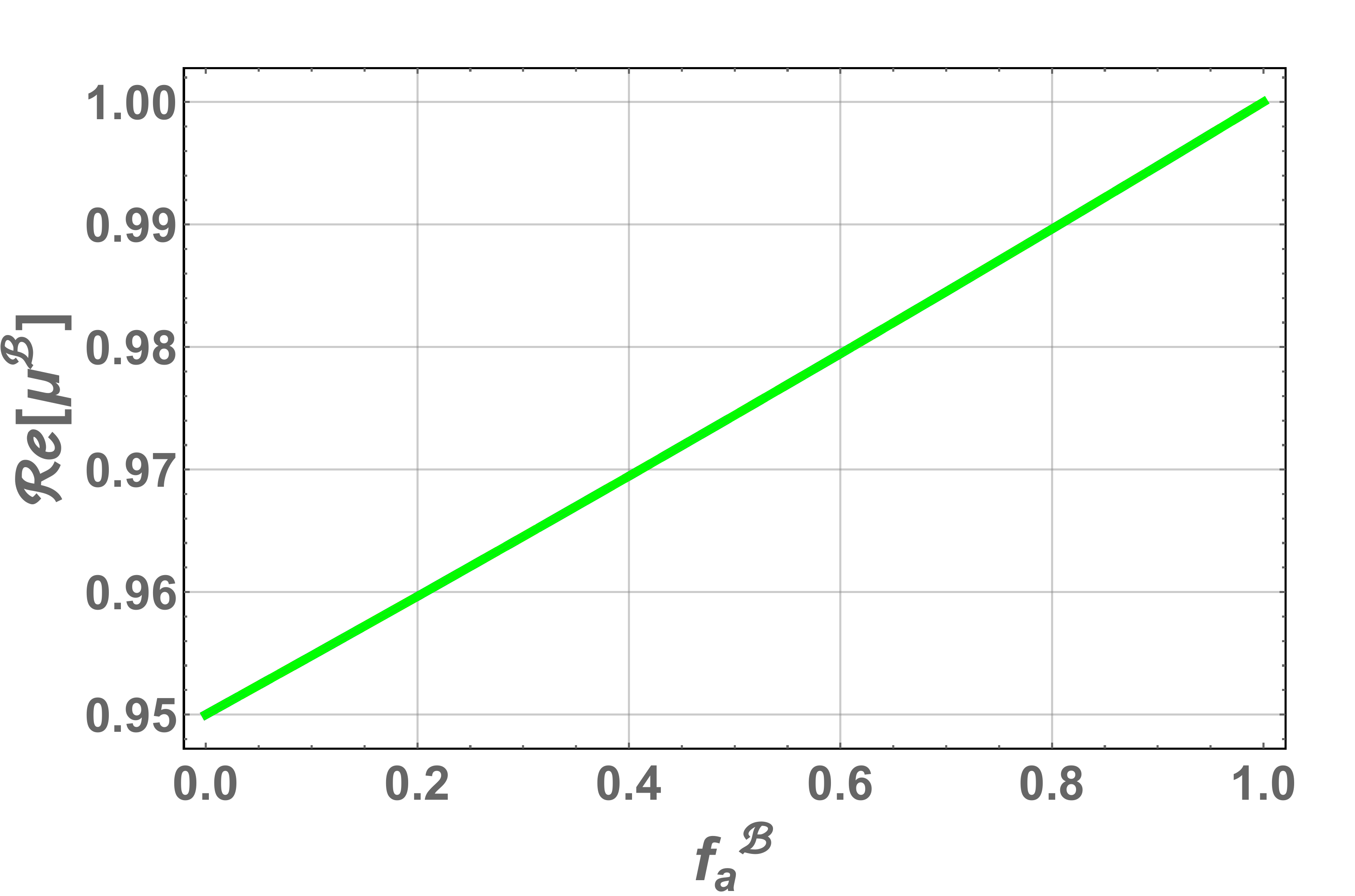} \includegraphics[width=5.8cm]{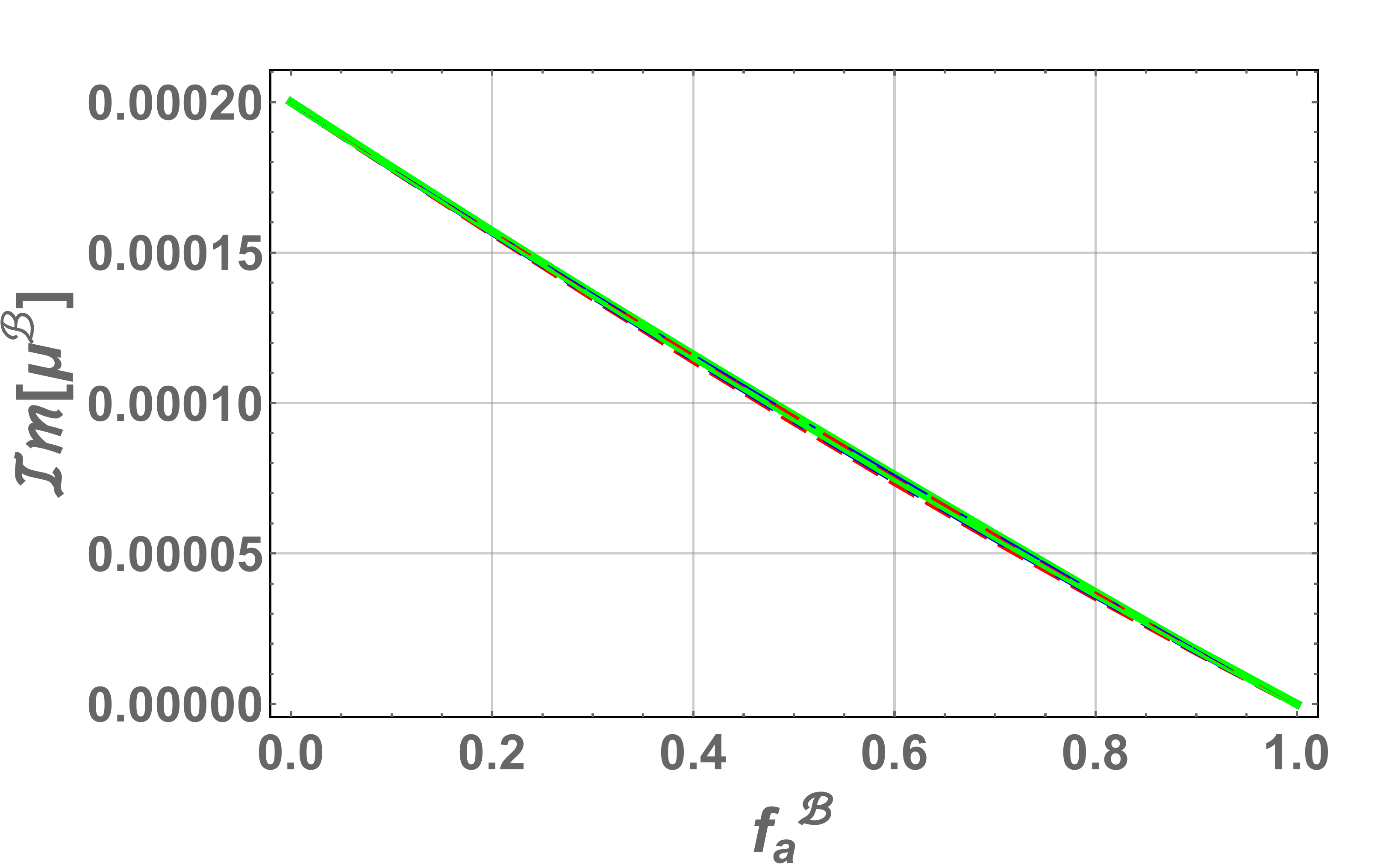} \\
\includegraphics[width=5.8cm]{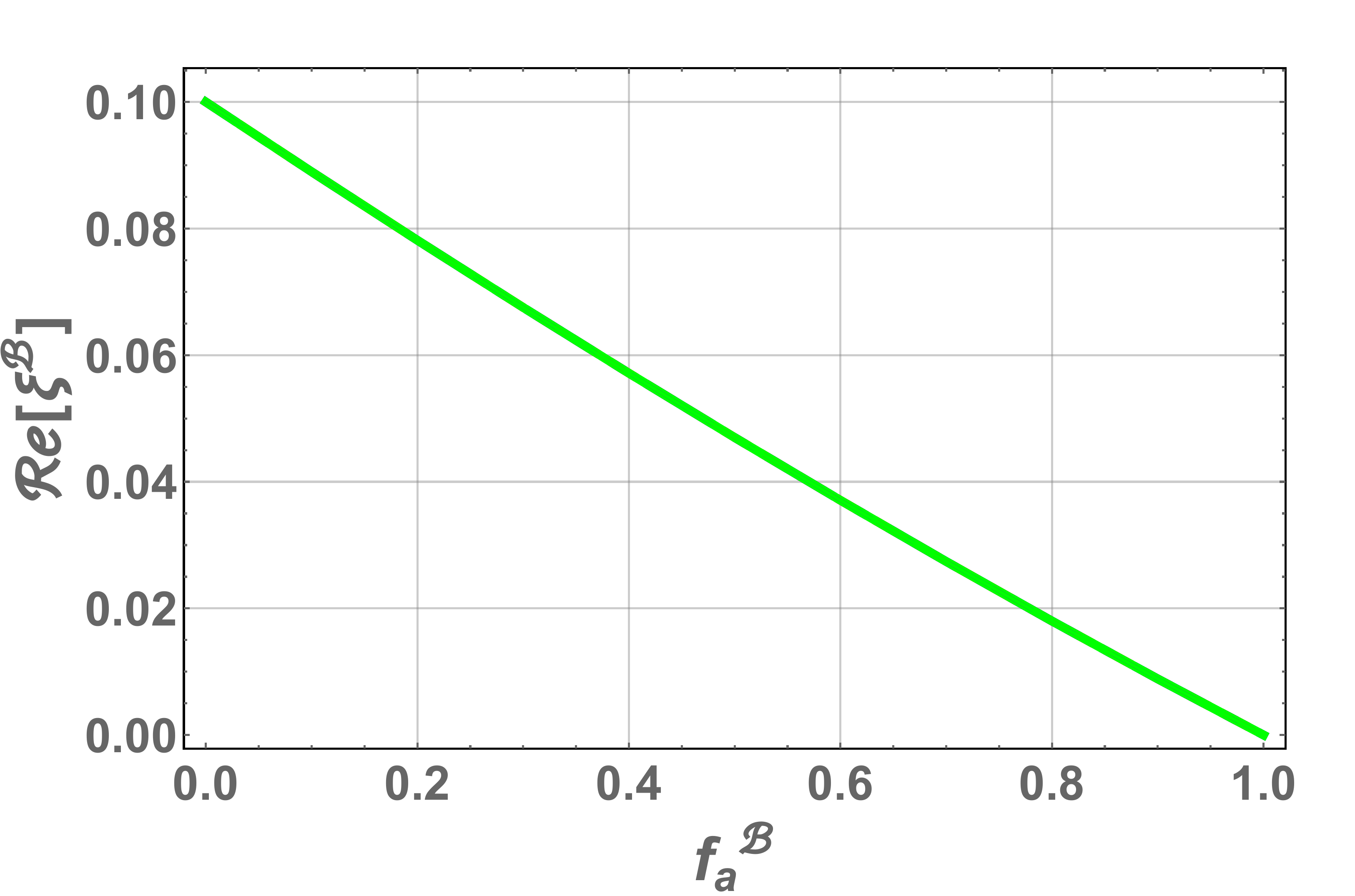} \includegraphics[width=5.8cm]{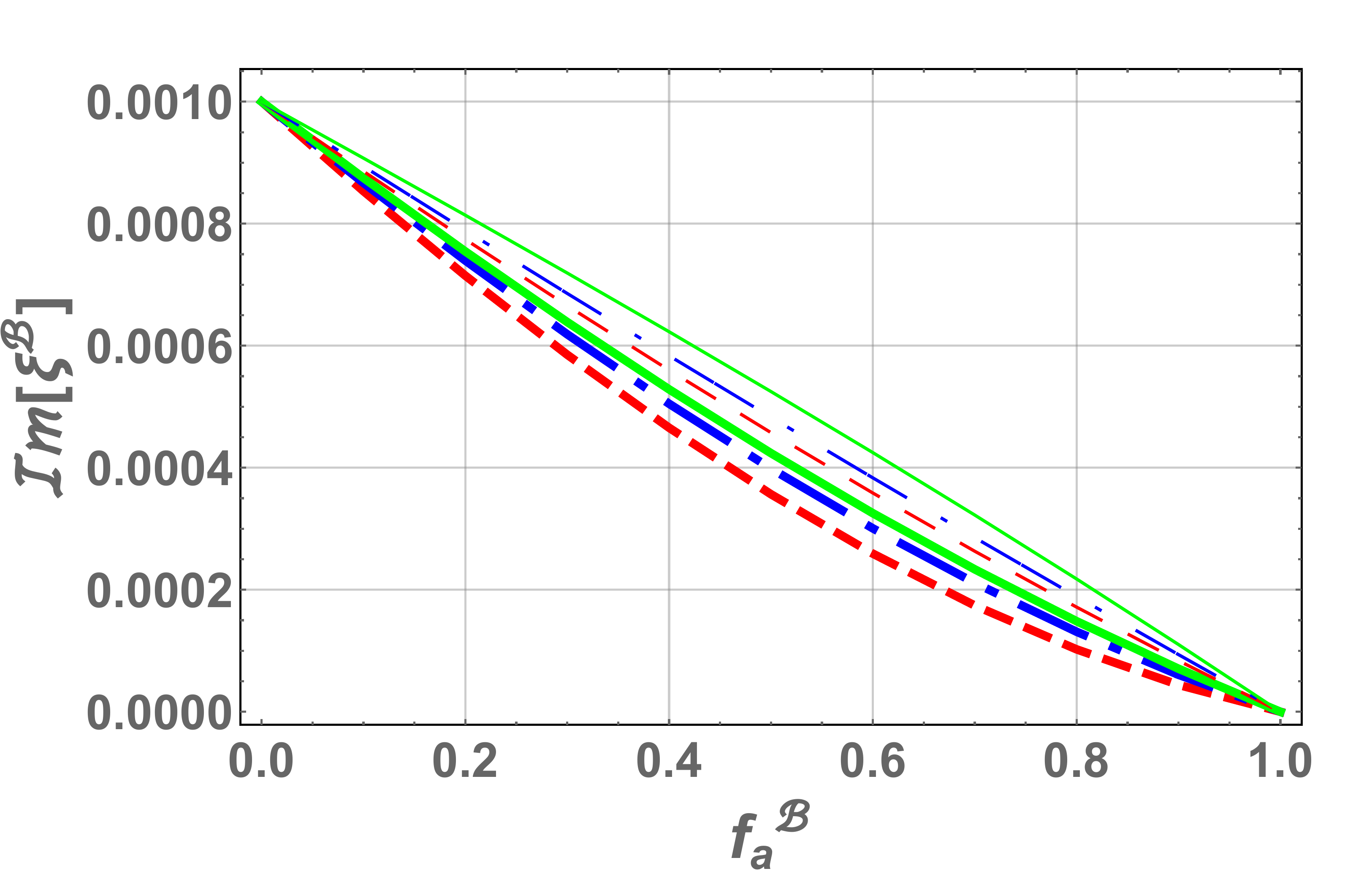}
 \caption{Bruggeman estimates of the real and imaginary parts of the relative constitutive parameters $\eps^\calB$, $\mu^\calB$, and $\xi^\calB$ plotted against volume fraction $f^\calB_a$ for $\eps^\calB_a = 2 - 0.02 d  i $. Key: $d= 1.0 $ (thick dashed red curve), 0.5 (thick broken dashed blue curve), 0.2 (thick solid green curve), $-0.2$ (thin dashed red curve), $-0.5$ (thin broken dashed blue curve), and $-1.0$ (thin solid green curve).
 } \label{fig1}
\end{figure}

\begin{figure}[!htb]
\centering
\includegraphics[width=5.8cm]{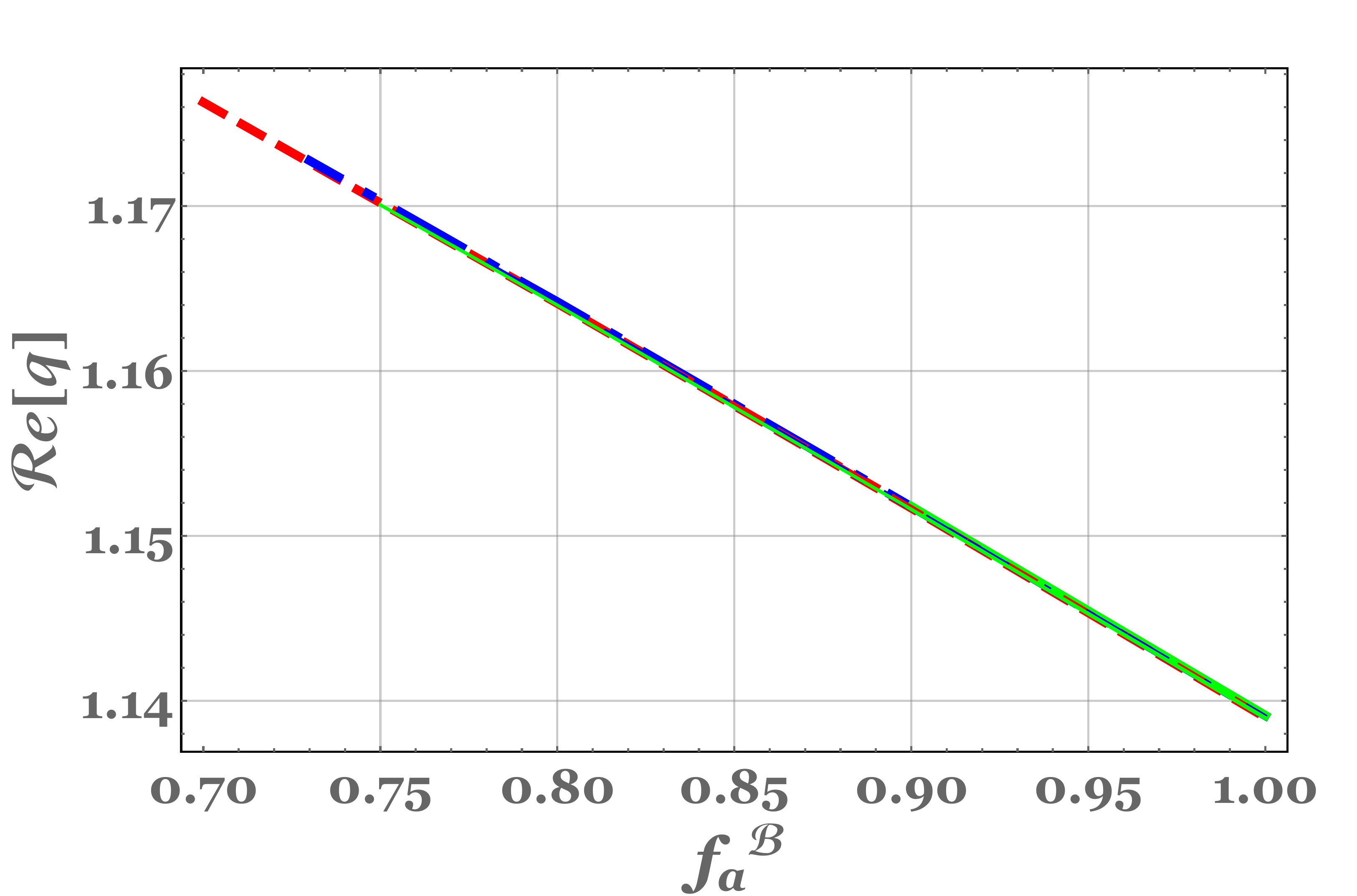}
\includegraphics[width=5.8cm]{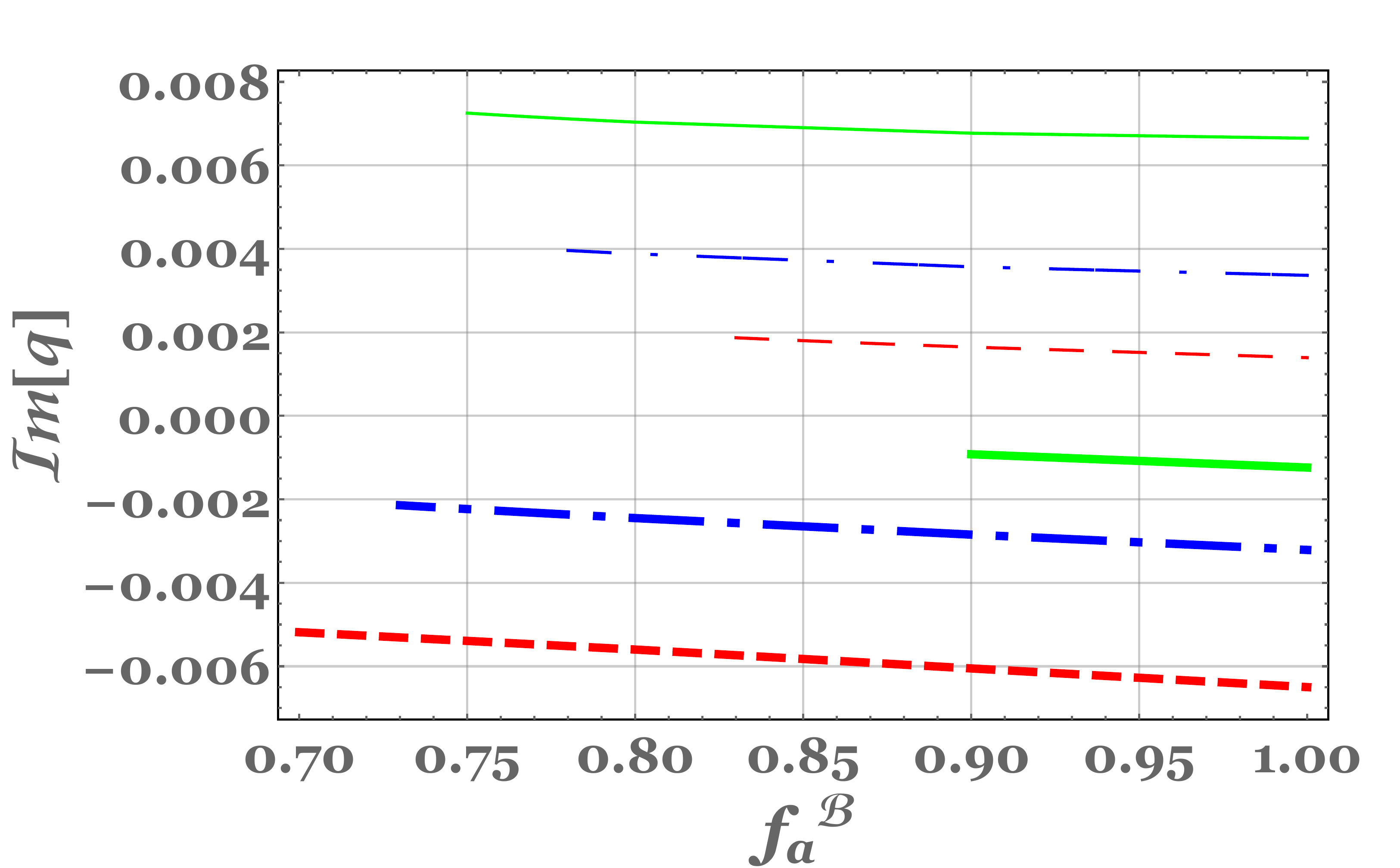}
 \caption{Real and imaginary parts of the relative wavenumber $q$ plotted against volume fraction $f^\calB_a$ for $\eps^\calB_a = 2 - 0.02 d  i $. Key: $d= 1.0 $ (thick dashed red curve), 0.5 (thick broken dashed blue curve), 0.2 (thick solid green curve), $-0.2$ (thin dashed red curve), $-0.5$ (thin broken dashed blue curve), and $-1.0$ (thin solid green curve).
 } \label{fig2}
\end{figure}

\begin{figure}[!htb]
\centering
\includegraphics[width=5.8cm]{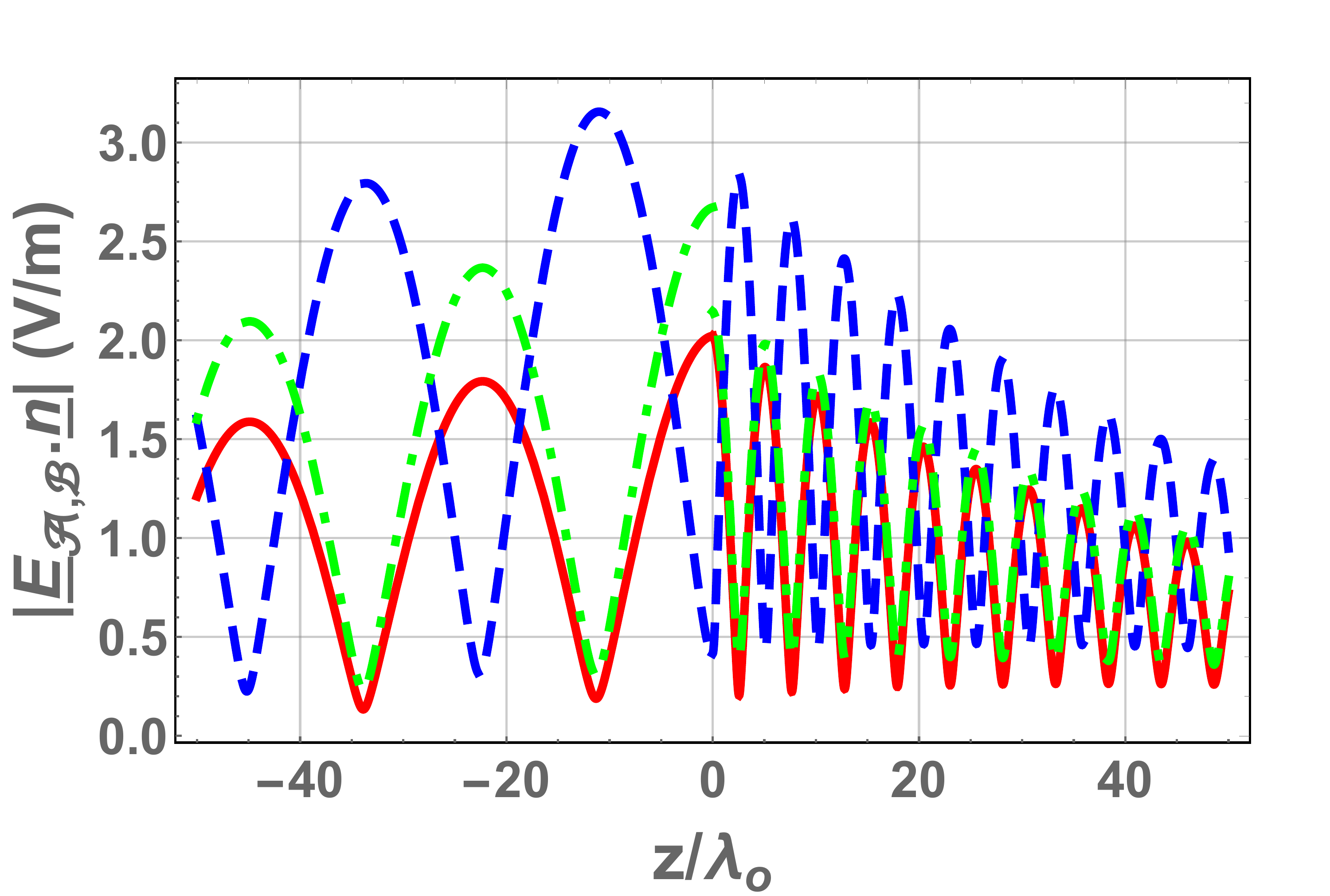}
\includegraphics[width=5.8cm]{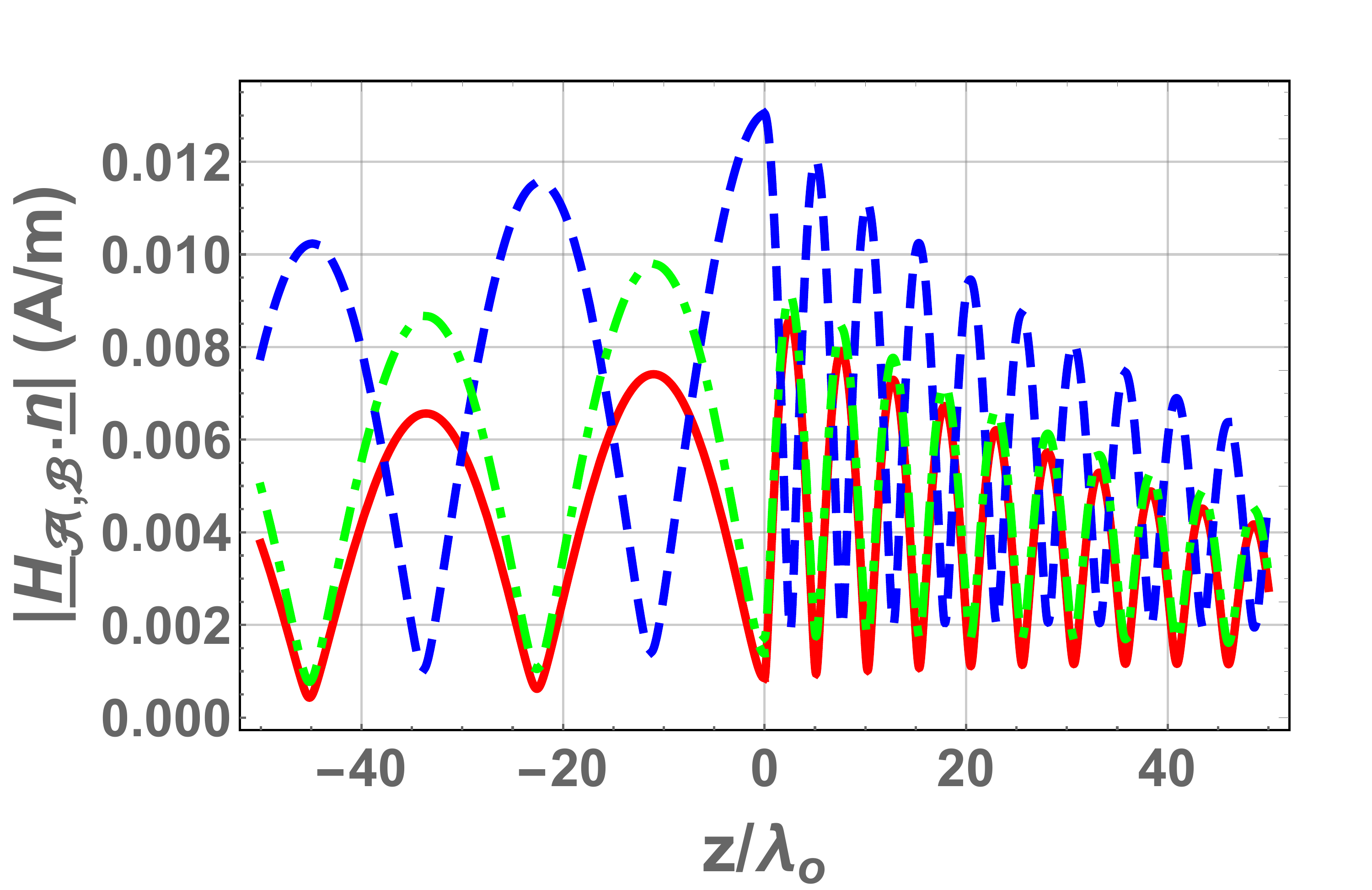}\\
\includegraphics[width=5.8cm]{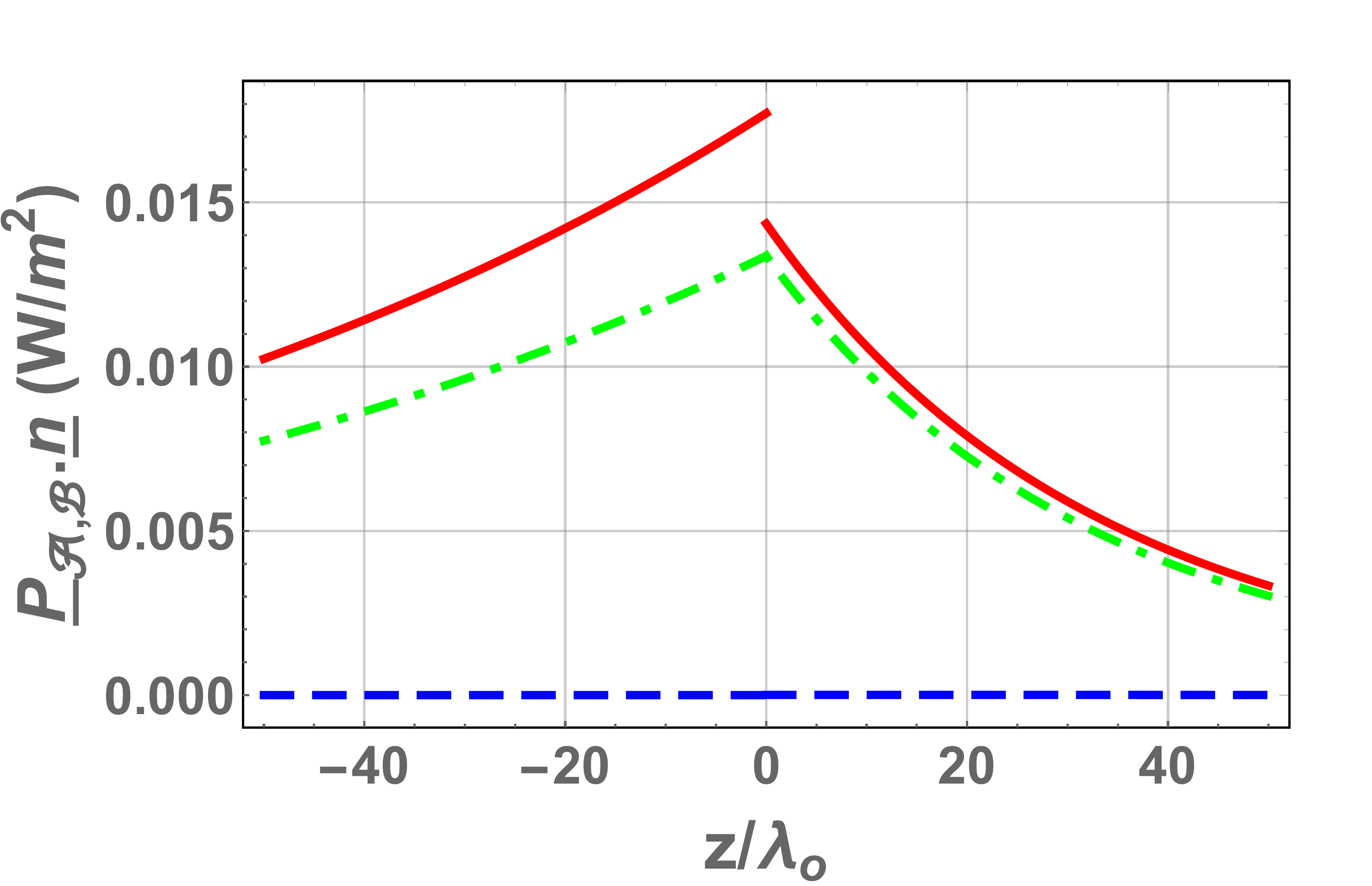}
 \caption{Magnitudes of $\underline{E}_{\, \mathcal{A},\mathcal{B}} (z\hat{\underline{u}}_{\,z}) \. \#n$, and $\underline{H}_{\, \mathcal{A},\mathcal{B}}  (z\hat{\underline{u}}_{\,z}) \. \#n$, along with $\underline{P}_{\, \mathcal{A},\mathcal{B}}  (z\hat{\underline{u}}_{\,z}) \. \#n$, plotted versus $z/\lambdao$,
 when   $\eps^\calB_a = 2 - 0.02 d  i $, $f^\calB_a = 0.85$, $d = 0.5$, and $C_{\mathcal{A}1} = 1$ V m${}^{-1}$.
 Key:   $\#n = \ux$ solid red curves; $\#n = \uy$ dashed blue curves; $\#n = \uz$ broken dashed curves.
 } \label{fig3}
\end{figure}

\begin{figure}[!htb]
\centering
\includegraphics[width=5.8cm]{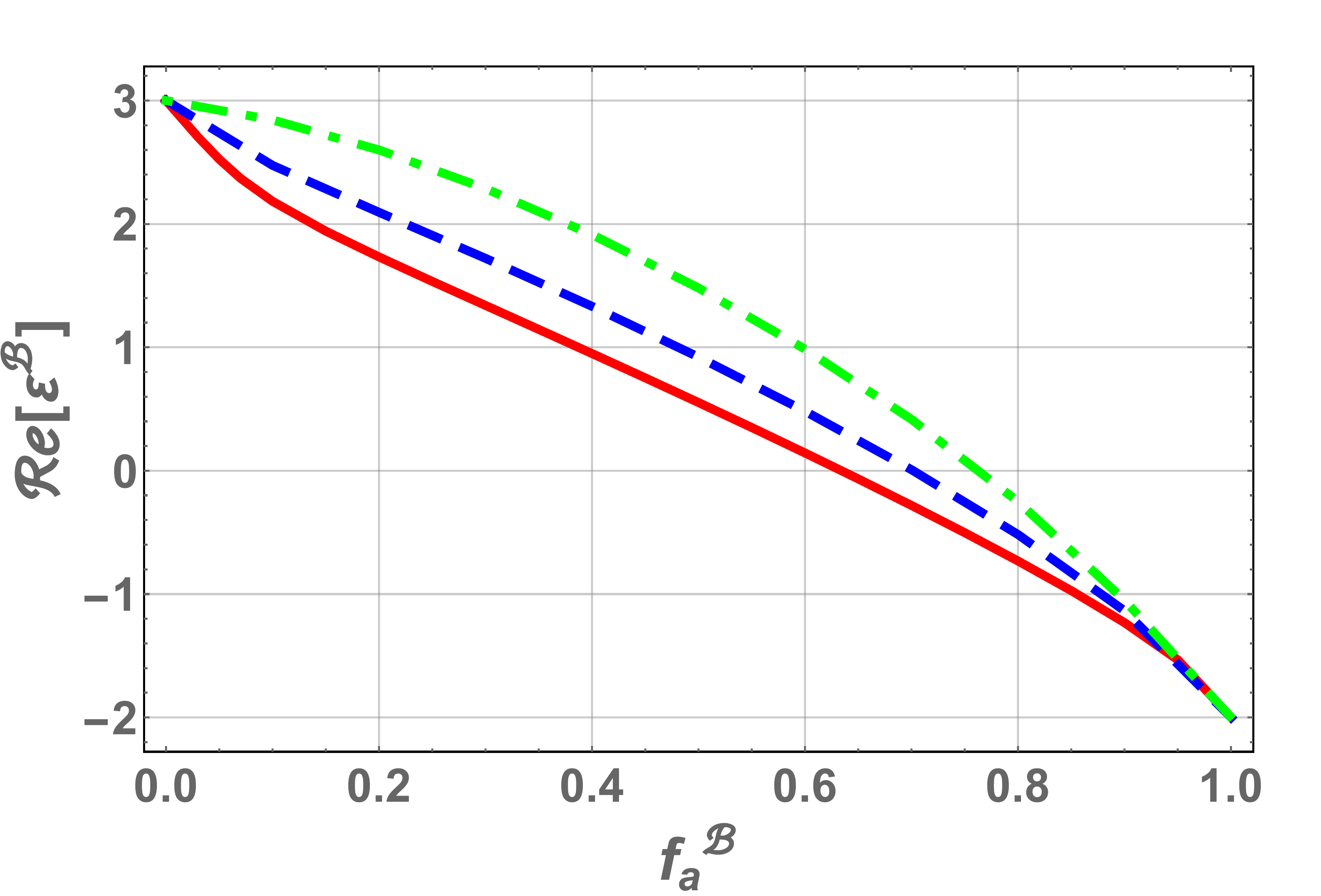} \includegraphics[width=5.8cm]{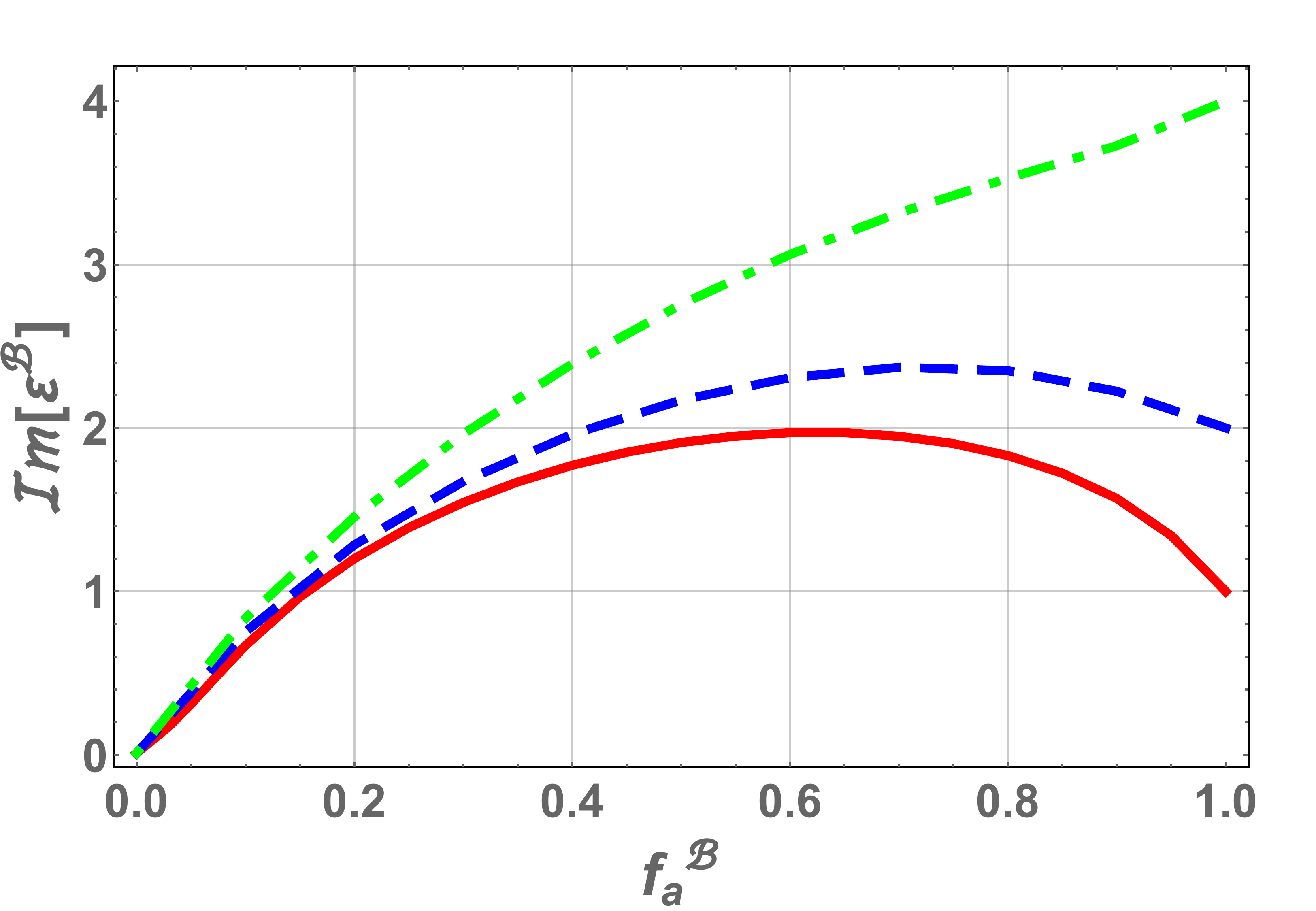}\\
\includegraphics[width=5.8cm]{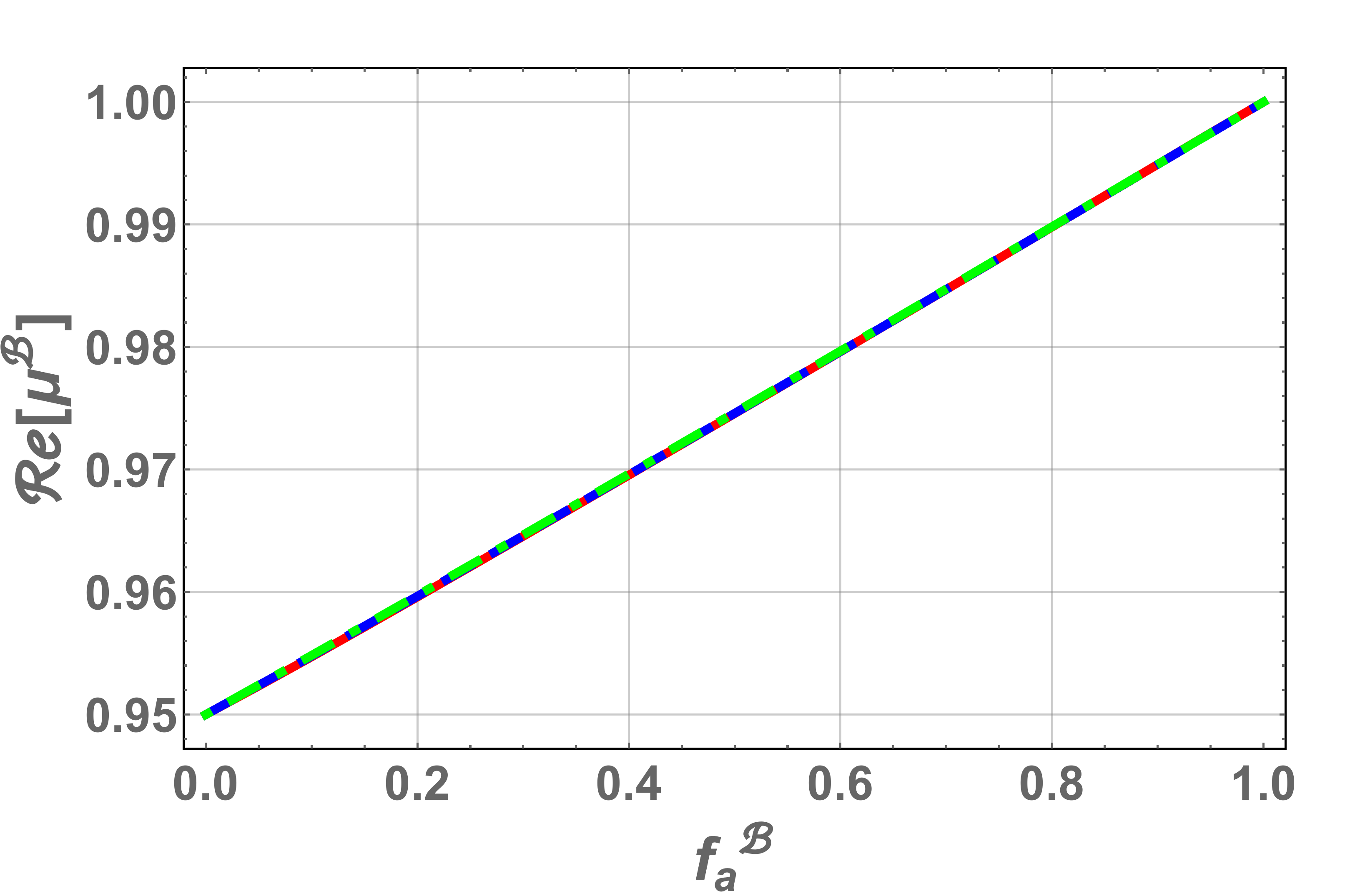} \includegraphics[width=5.8cm]{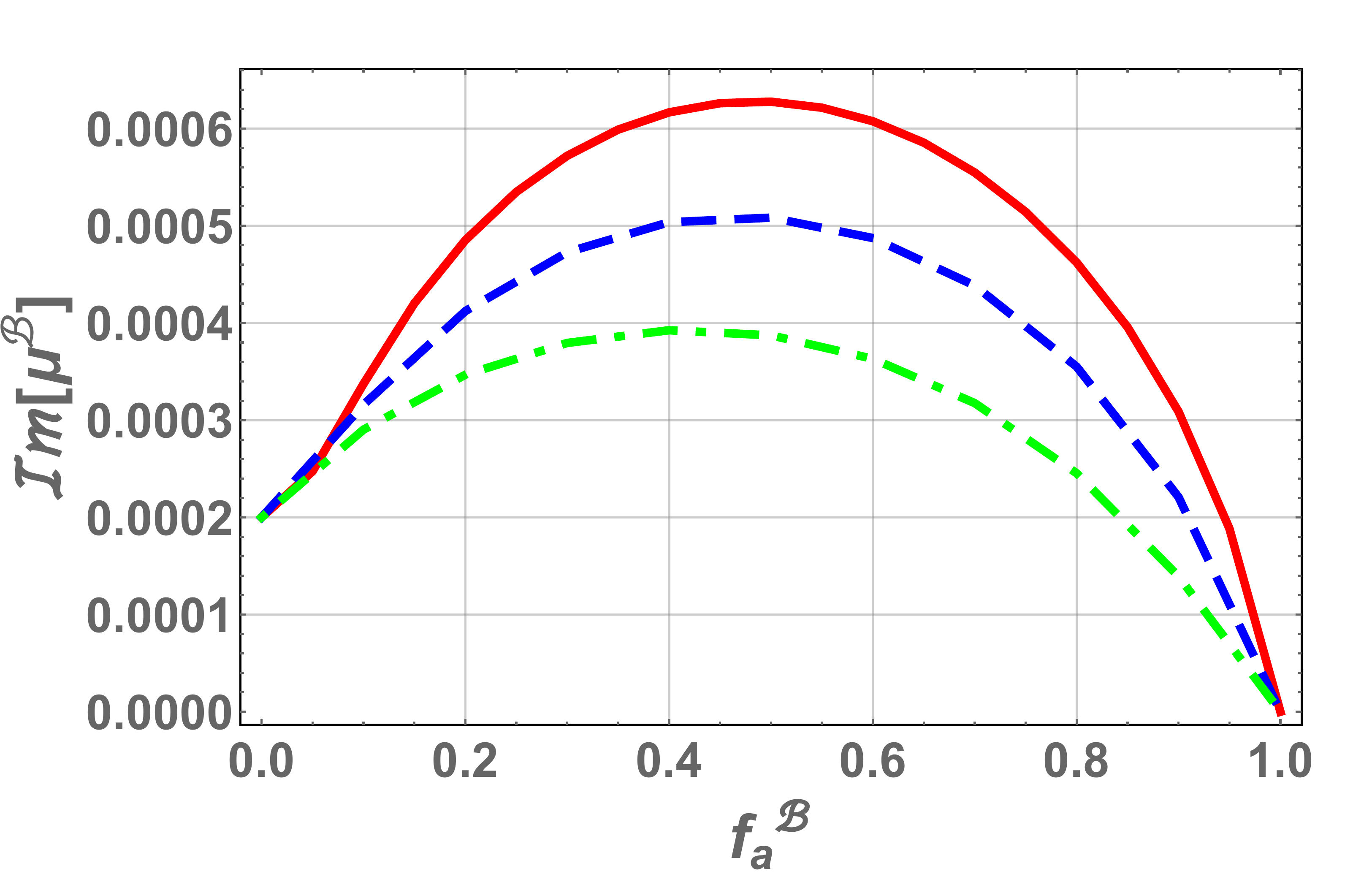} \\
\includegraphics[width=5.8cm]{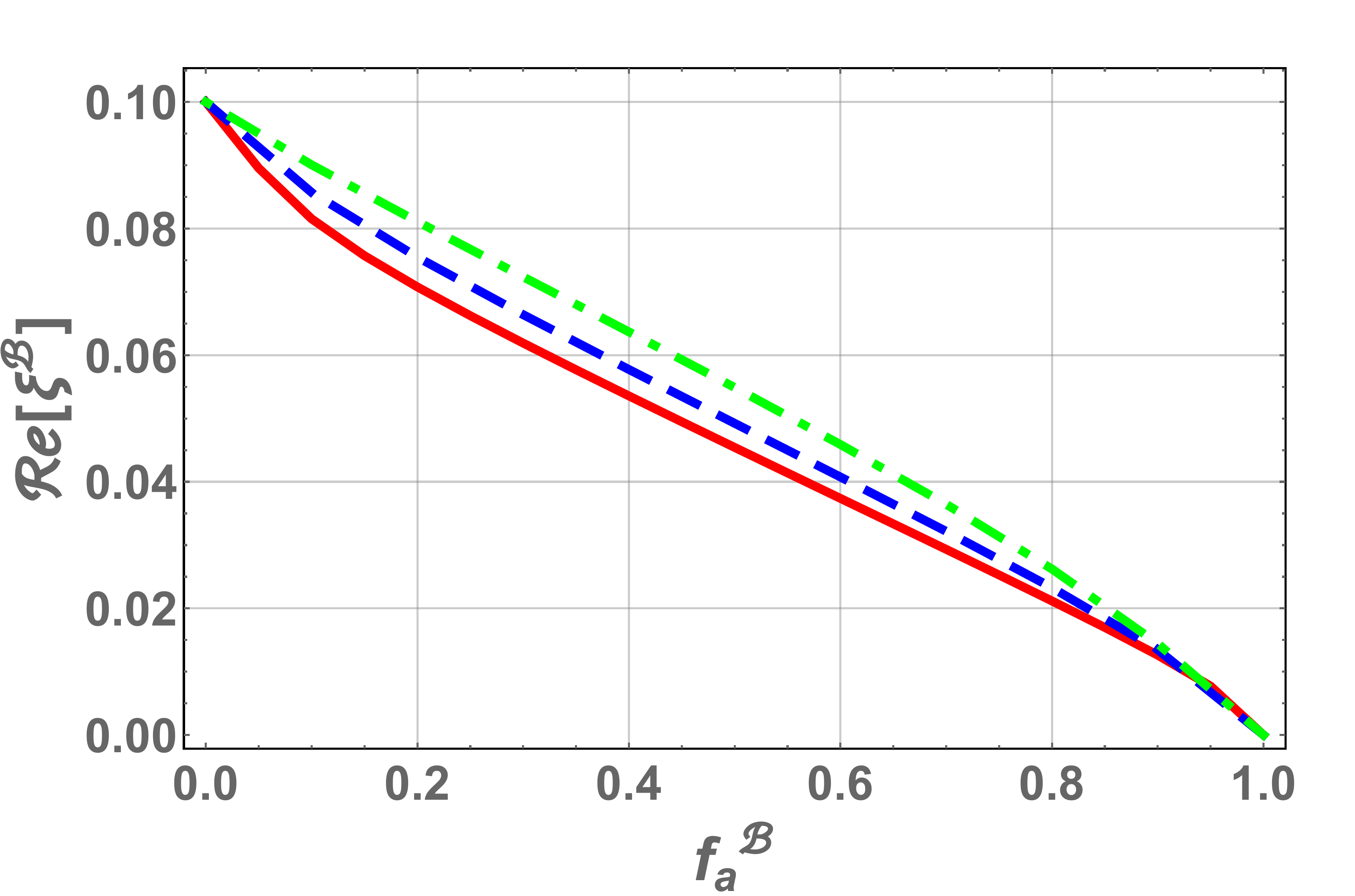} \includegraphics[width=5.8cm]{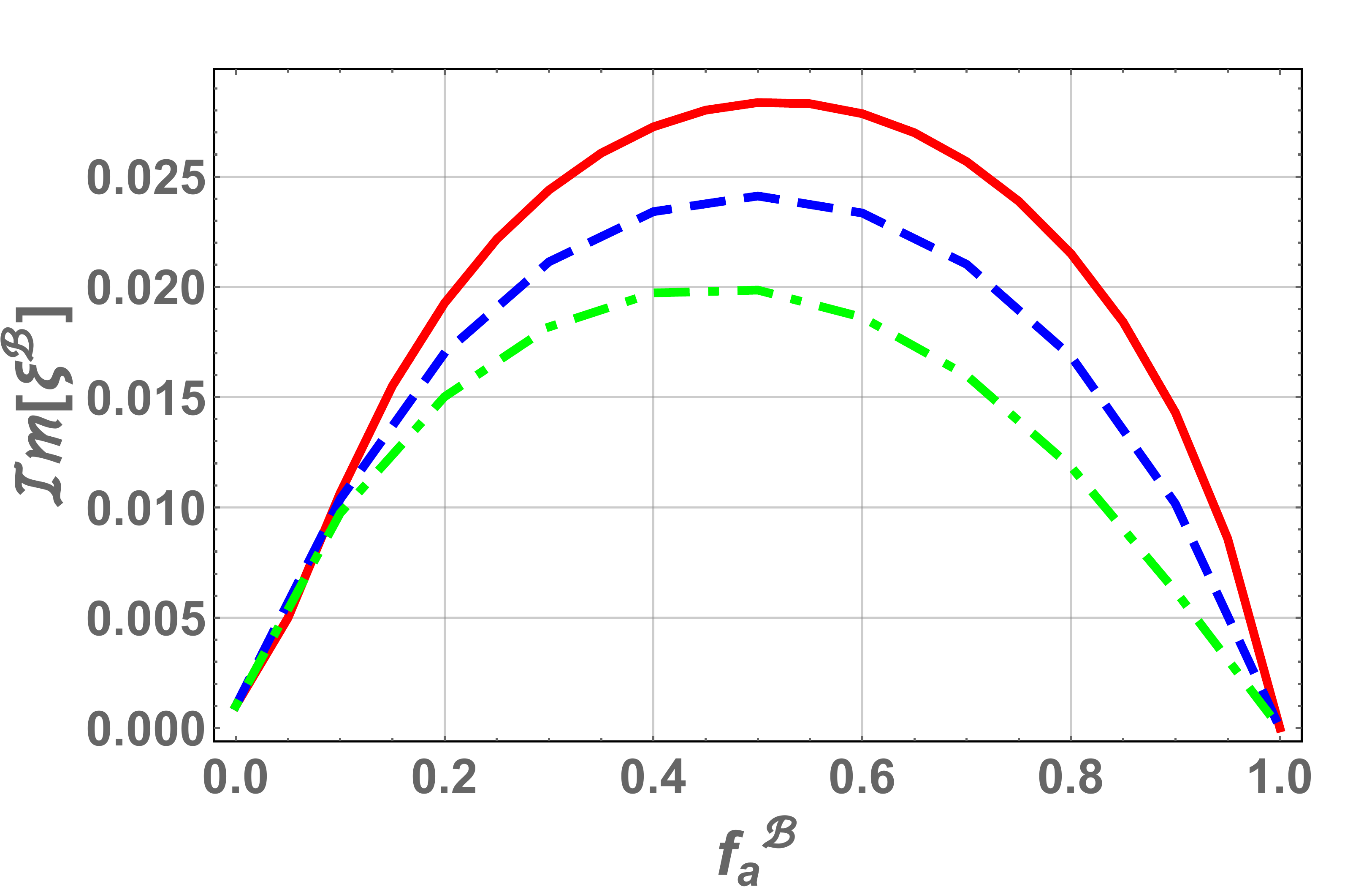}
 \caption{Bruggeman estimates of the real and imaginary parts of the relative constitutive parameters $\eps^\calB$, $\mu^\calB$, and $\xi^\calB$ plotted against volume fraction $f^\calB_a$ for $\eps^\calB_a = -2 + 0.02 d  i $. Key: $d= 50 $ (solid red curve), 100 (dashed blue curve), and 200 (broken dashed green curve).
 } \label{fig4}
\end{figure}

\begin{figure}[!htb]
\centering
\includegraphics[width=5.8cm]{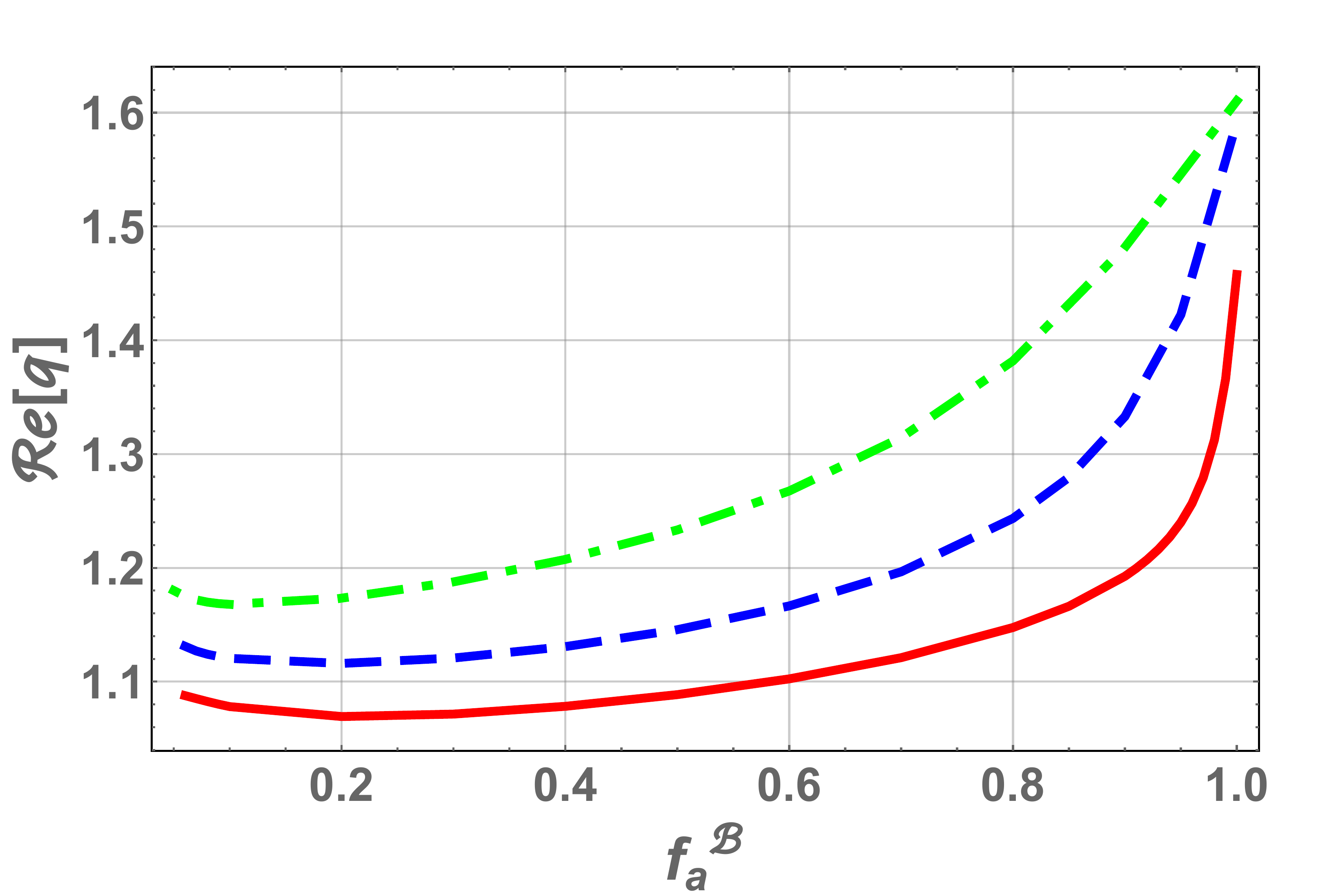}
\includegraphics[width=5.8cm]{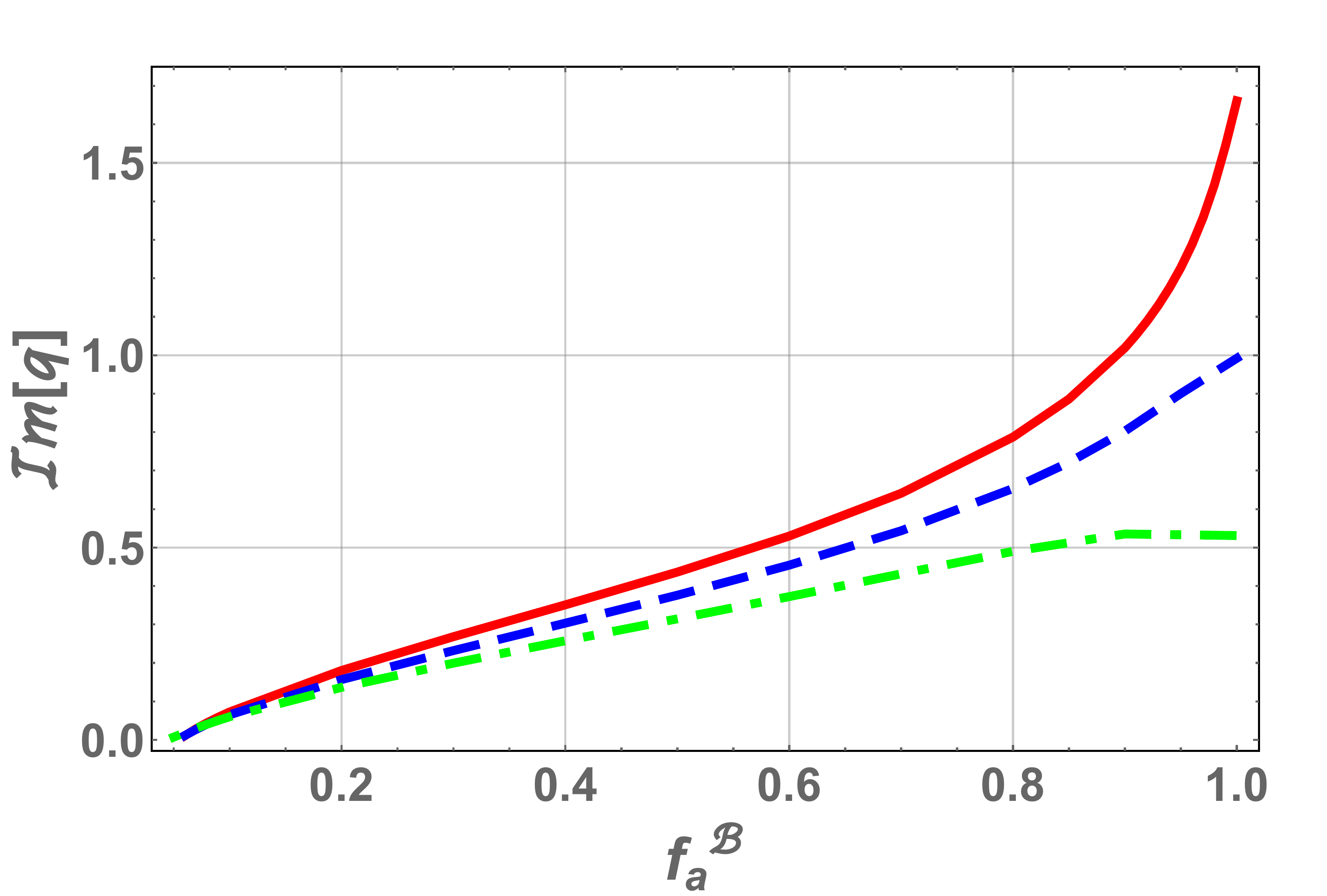}
 \caption{Real and imaginary parts of the relative wavenumber $q$ plotted against volume fraction $f^\calB_a$
 for $\eps^\calB_a = -2 + 0.02 d  i $. Key: $d= 50 $ (solid red curve), 100 (dashed blue curve), and 200 (broken dashed green curve).
 } \label{fig5}
\end{figure}

\begin{figure}[!htb]
\centering
\includegraphics[width=5.8cm]{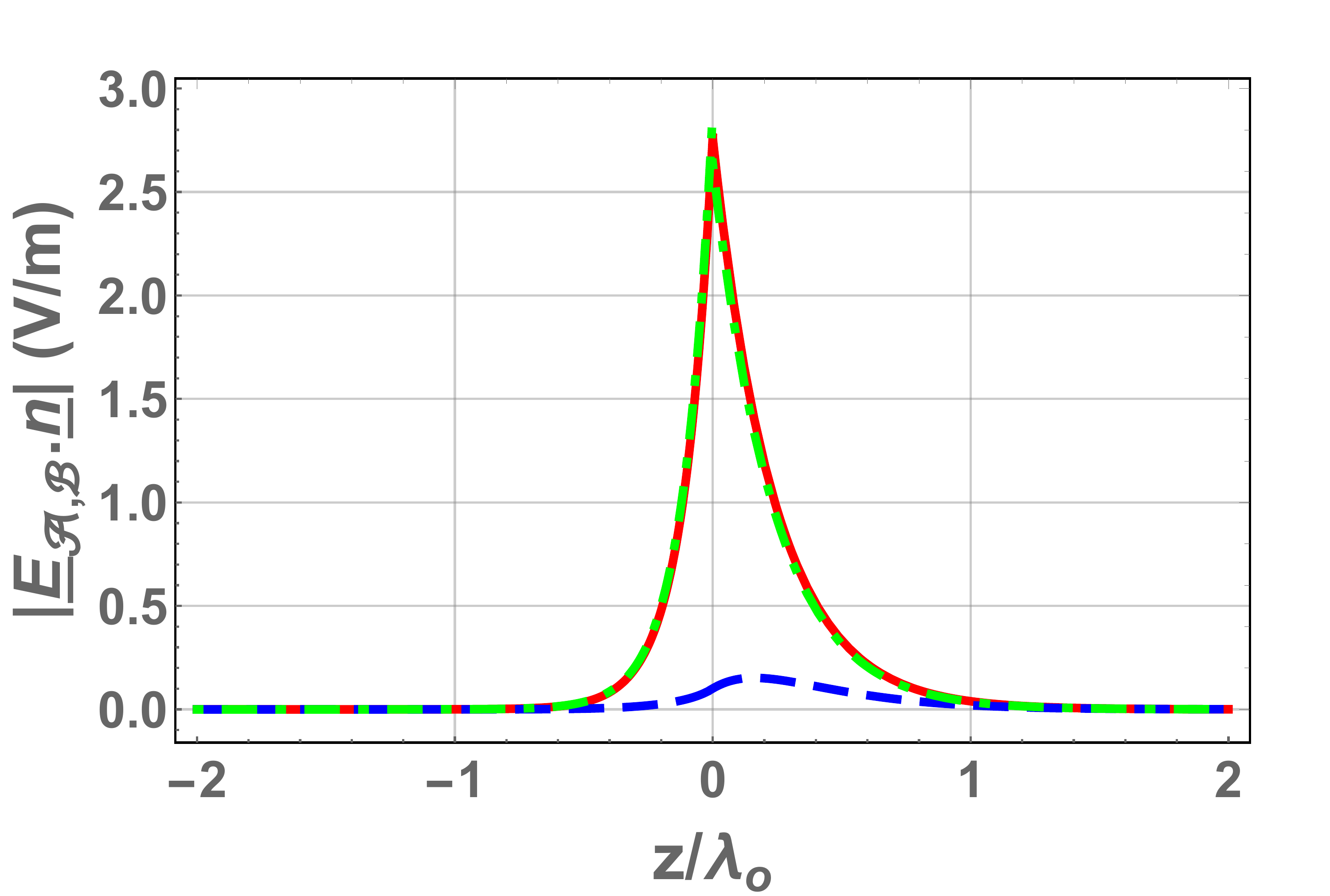}
\includegraphics[width=5.8cm]{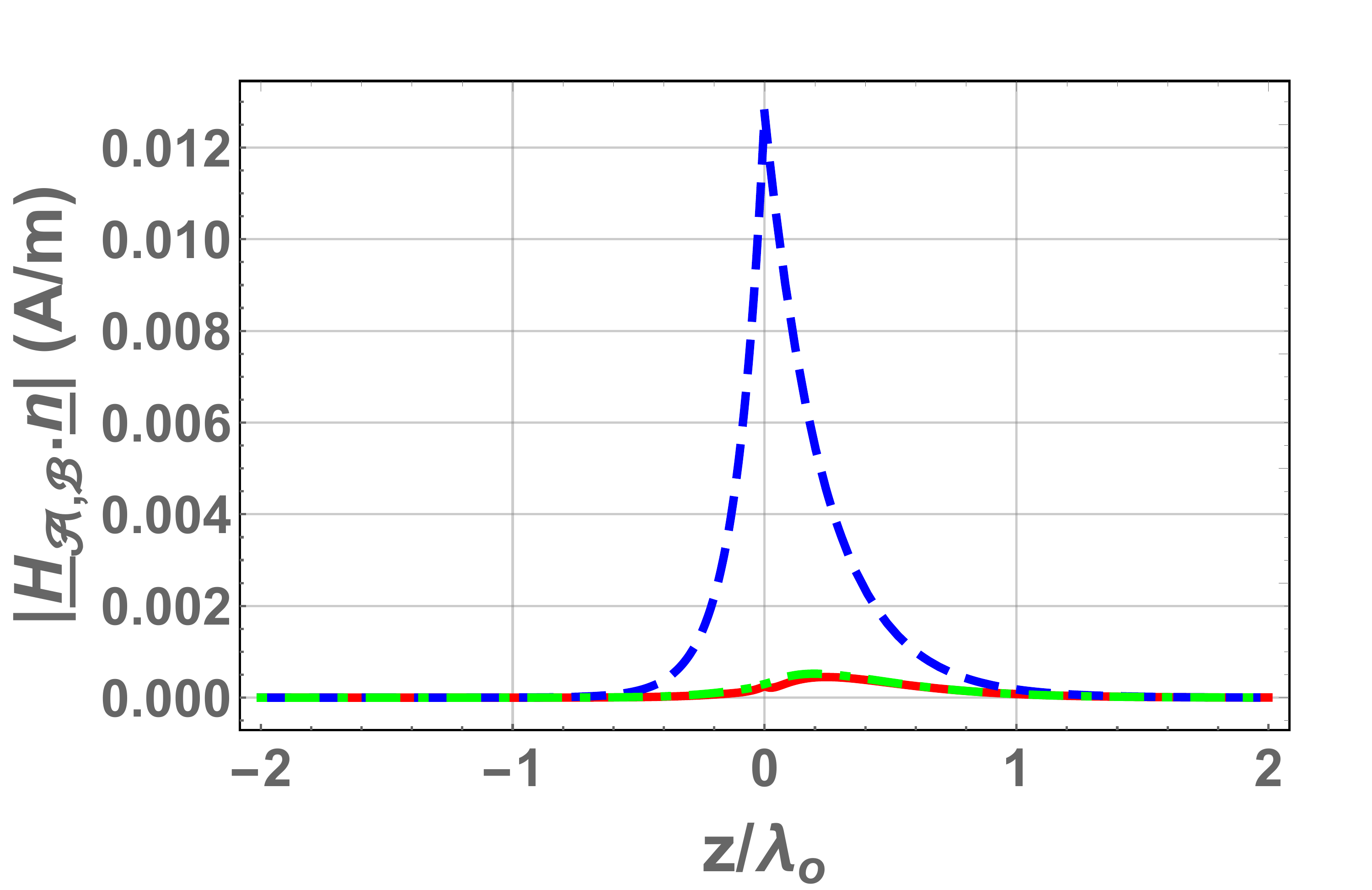}\\
\includegraphics[width=5.8cm]{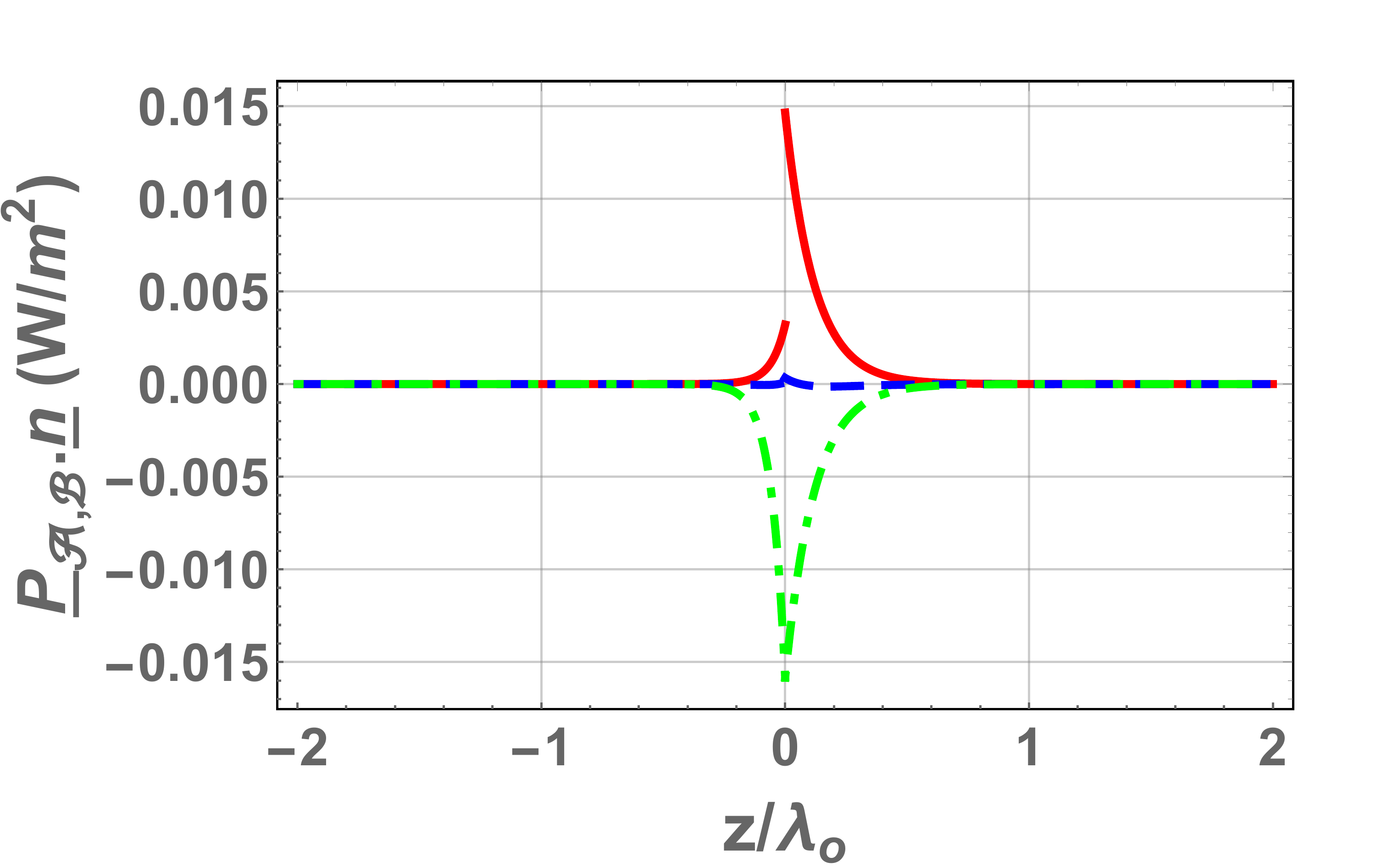}
 \caption{Magnitudes of $\underline{E}_{\, \mathcal{A},\mathcal{B}} (z\hat{\underline{u}}_{\,z}) \. \#n$, and $\underline{H}_{\, \mathcal{A},\mathcal{B}}  (z\hat{\underline{u}}_{\,z}) \. \#n$, along with $\underline{P}_{\, \mathcal{A},\mathcal{B}}  (z\hat{\underline{u}}_{\,z}) \. \#n$, plotted versus $z/\lambdao$,
 when  $\eps^\calB_a = -2 + 0.02 d  i $, $f^\calB_a = 0.85$, $d = 100$, and $C_{\mathcal{A}1} = 1$ V m${}^{-1}$.
 Key:   $\#n = \ux$ solid red curves; $\#n = \uy$ dashed blue curves; $\#n = \uz$ broken dashed curves.
 } \label{fig6}
\end{figure}

We begin our presentation of numerical results with the case where the achiral component material $\calB_a$ is 
specified by the relative permittivity $\eps^\calB_a = 2 - 0.02 d  i $. Let us consider  $d \in \lec  -0.2, -0.5, -1 \ric$ for which component material $\calB_a$ is 
 a weakly  dissipative  dielectric material and  $d \in \lec 0.2, 0.5, 1 \ric$ for which component material $\calB_a$ is 
 an active  dielectric material. Estimates of the relative constitutive parameters of partnering material $\calB$, as provided by the Bruggeman equation \r{Br_eq}, are  plotted as functions of volume fraction $f^\calB_a$ in Fig.~\ref{fig1}.  
The real parts of  $\eps^\calB$, $\mu^\calB$, and $\xi^\calB$ are almost independent of the parameter $d$; 
they vary in an approximately linear manner as $f^\calB_a$ varies. The imaginary part of  $\eps^\calB$ is negative valued for $d>0$ and positive valued for $d<0$. And the magnitude  $\left| \mbox{Im} \lec \eps^\calB \ric \right|$ is larger when the magnitude of $d$ is larger. The imaginary parts of $\mu^\calB$ and $\xi^\calB$ are both
much less sensitive to $d$ than is  $ \mbox{Im} \lec \eps^\calB \ric $; both are positive valued for all values of $d$ and  both decay to zero in the limit $f^\calB_a \to 1$.

 For  $d \in \lec  -1, -0.5, -0.2, 0.2, 0.5, 1 \ric$, a surface wave is supported for certain ranges of volume fraction $f^\calB_a$. The real and imaginary parts of the relative wavenumber $q$ for these surface waves are plotted against $f^\calB_a$ in Fig.~\ref{fig2}. The volume fraction ranges of  that support surface waves are as follows:
 $f^\calB_a \in \le 0.75, 1 \ris$ for $d = -1$, 
$f^\calB_a \in \le 0.78, 1 \ris$ for $d = -0.5$,
$f^\calB_a \in \le 0.83, 1 \ris$ for $d = -0.2$,  
$f^\calB_a \in \le 0.90, 1 \ris$ for $d = 0.2$, 
$f^\calB_a \in \le 0.73, 1 \ris$ for $d = 0.5$, 
and
$f^\calB_a \in \le 0.70, 1 \ris$ for $d = 1$.
Notably, surface waves are not supported at all for small values of  $f^\calB_a $.
The real parts of $q$ decrease approximately linearly as $f^\calB_a$ increases, and these values are almost independent of $d$.
The imaginary parts of $q$ decrease slightly as $f^\calB_a$ increases, with the magnitude $\left| \mbox{Im} \lec q \ric \right|$ being greater when the magnitude of $d$ is greater. Also $ \mbox{Im} \lec q \ric >0 $ when $d<0$ and
$ \mbox{Im} \lec q \ric < 0 $ when $d>0$; that is, the surface wave attenuates in the direction of propagation when the component material $\calB_a$ is dissipative and
the surface wave is amplified in the direction of propagation when the component material $\calB_a$ is active.

Further illumination on the nature of these surface waves is offered in   Fig.~\ref{fig3} wherein profiles of the electric and magnetic field  phasors are
provided. Specifically, plotted are    $|\underline{E}_{\, \ell} (z\uz) \. \#n|$
and $|\underline{H}_{\, \ell} (z\uz) \. \#n|$,
$\ell \in \lec \mathcal{A}, \mathcal{B} \ric$,
 versus $z / \lambdao$ for $\#n \in \lec \ux, \uy, \uz \ric$,
 when $d=0.5$ and
  $f^\calB_a = 0.85$, with $C_{\mathcal{A}1} = 1$ V m${}^{-1}$.
  Also plotted are the corresponding profiles of the Cartesian components 
$\underline{P}_{\, \ell} (z\uz) \. \#n$, $\ell \in \lec \mathcal{A}, \mathcal{B} \ric$ and $\#n \in \lec \ux, \uy, \uz \ric$, of the time-averaged Poynting vector \c{Chen}
\begin{equation}
\underline{P}_{\, \ell} (\#r) = \frac{1}{2} \mbox{Re} \les \, \underline{E}_{\, \ell} (\#r)  \times 
\underline{H}^*_{\, \ell} (\#r) \, \ris, \qquad \ell \in \lec  \mathcal{A}, \mathcal{B}  \ric\,,
\end{equation}
where the asterisk denotes the complex conjugate.
The surface wave is rather loosely localized to the  planar  interface $z=0$, with 
 significant spreading of the fields into
 both the half-spaces $z>0$ and $z<0$ even at $z = \pm 40 \lambdao$. The   surface wave is a little more tightly bound to the planar interface
 $z=0$ in the half-space $z>0$ than in the half-space $z<0$.
Profiles of the field phasors are qualitatively similar for the other values of $d$ considered here.

Next let us explore the case where the achiral component material $\calB_a$ is a plasmonic material. To this end, we select
 the relative permittivity $\eps^\calB_a = - 2 + 0.02 d  i $ with $d \in \lec 50, 100, 200 \ric$.
In Fig.~\ref{fig4} estimates of the relative constitutive parameters of partnering material $\calB$, as provided by the Bruggeman equation \r{Br_eq}, are  plotted as functions of volume fraction $f^\calB_a$.  
The real parts of  $\eps^\calB$ and $\xi^\calB$ are generally larger when  the parameter $d$ is larger, especially for mid-range values of $f^\calB_a$, whereas 
the real part of   $\mu^\calB$ is almost independent of $d$.
 The imaginary part of  $\eps^\calB$ is larger for larger values of $d$. In contrast, 
the imaginary parts of  $\mu^\calB$ and $\xi^\calB$ are larger for smaller values of $d$
and these  quantities both decay to zero in the limit $f^\calB_a \to 1$.

With the plasmonic component material $\calB_a$ and $d \in \lec  50, 100, 200 \ric$, a surface wave is supported for  wide 
 ranges of volume fraction $f^\calB_a$, but not for all values of $f^\calB_a$. The real and imaginary parts of the relative wavenumber $q$ for these surface waves are plotted against $f^\calB_a$ in Fig.~\ref{fig5}. Surface waves are supported for the following ranges of volume fraction:
 $f^\calB_a \in \le 0.06, 1 \ris$ for $d = 50$, 
$f^\calB_a \in \le 0.06, 1 \ris$ for $d = 100$, and
$f^\calB_a \in \le 0.05, 1 \ris$ for $d = 200$.
Notably, surface waves are not supported in the limit $f^\calB_a \to 0$.
The real parts of $q$ increase quite sharply as $f^\calB_a$ increases, depending upon the value $d$.
Likewise, the imaginary parts of $q$ increase  as $f^\calB_a$ increases, with the largest values of  $\mbox{Im} \lec q \ric $ arising when the magnitude of $d$ is smallest. Also, for all values of $d$,  $ \mbox{Im} \lec q \ric >0 $ which indicates  that the surface waves are attenuated in  the direction of propagation.

Profiles of the electric and magnetic field phasors are
presented in   Fig.~\ref{fig6},  for the case where
component material $\calB_a$ is a plasmonic material,
 in order to shed further light on
the nature of these surface waves. For these computations, 
 $d=100$ and
  $f^\calB_a = 0.85$, with $C_{\mathcal{A}1} = 1$ V m${}^{-1}$.
  Also plotted are the corresponding profiles of the Cartesian components  of the time-averaged Poynting vector.
The surface wave is quite tightly localized to the  planar  interface $z=0$, much more so than the corresponding surface wave represented in Fig.~\ref{fig3}, with 
 the surface wave being essential confined to the region $ - \lambdao < z< \lambdao$. The   surface wave is rather more tightly bound to the planar interface
 $z=0$ in the half-space $z<0$ than in the half-space $z>0$.
Profiles of the field phasors are qualitatively similar for the other values of $d$ considered here.


\section{Closing remarks}

The planar interface of two isotropic chiral materials has been shown to support surface-wave propagation  for certain constitutive parameter ranges.
When the component material $\calB_a$ is a plasmonic material  the surface waves are akin to SPP waves. 
But when the component material $\calB_a$ is a dielectric material  the surface wave is not akin to a SPP wave, since $\mbox{Re} \lec \eps^\calA \ric > 0$
and  $\mbox{Re} \lec \eps^\calB \ric > 0$. Furthermore when the component material $\calB_a$ is a dielectric material  the surface wave is not akin to a Dyakonov wave, since both partnering materials are isotropic. These  surface waves are not akin to any of the well-established types  of surface wave \c{ESW_book}.

Notably, for all surface-wave solutions reported herein the constitutive parameters of the partnering materials are complex-valued with non-zero imaginary parts. The question arises: Is surface-wave propagation possible for  the idealized case in which the constitutive parameters of the partnering materials are real-valued?
 Owing to the intractability of the dispersion equation \r{disp_equation}, a definitive answer is not available.
However,  extensive numerical searches were undertaken in an attempt to find surface-wave solutions for real-valued constitutive parameters but none were found. 
Therefore, our  numerical evidence suggests that  the answer to this question is `no'.

\vspace{2mm}

\noindent {\bf Acknowledgments.}
MN acknowledges the support of the Higher Education Commission (HEC) of Pakistan for a six-month research visit to the University of Edinburgh via grant number HRD-2018-8668.

\end{document}